\documentstyle[psfig,aps]{revtex}
\newcounter{saveeqn}
\newcounter{App} 

\begin{document}
\hspace{11.4cm}hep-ph/0003254\\
\hspace*{11.8cm }LU TP 00-08\\
\vspace{0.5cm}

\thispagestyle{empty}
\begin{center}
{\Large\bf Semi-classical Description of Exclusive Meson Pair}\\
\vskip 0.5 true cm
{\Large \bf Production in $\gamma^*\gamma$ Scattering}
\vskip 1.5true cm

{\large\bf Martin Maul} 
\vskip 0.2true cm
{\it Institute for Theoretical Physics
Lund University, S\"olvegatan 14 a, S-223 62 Lund, Sweden} 
\end{center}
\vskip 4.0true cm
\begin{abstract}
\noindent
A semi-classical picture is given for the production
of exclusive meson pairs in  $\gamma^*\gamma$ scattering using elements
of the Lund string fragmentation model, spin and $C$-parity conservation.
The model can be generalized to the production of any few meson states
in scattering reactions at intermediate momentum transfers. As an 
example we show that we get a consistent description for the time-like
pion form factor. For the reaction $\gamma^*\gamma\to \pi^+\pi^-$ we find a
seizable cross section at LEP2 energies.
\end{abstract}
\vspace{3cm}
\newpage

\section{Introduction}
Recently, the factorization of the hadron production process
$\gamma^*\gamma \to h \bar h$ in a partonic handbag diagram and a 
two-hadron distribution amplitude  has been discussed 
\cite{Diehl:1998dk,Diehl:1998pg,Freund:2000xg}, which describes the exclusive 
fragmentation of a quark-antiquark pair into two hadrons.
This factorization is valid in the kinematic
region where the squared c.m. energy of the final hadrons $s = (p+p')^2$ 
is much smaller than the photon virtuality $Q^2$, see Fig.~\ref{kinema}.
So far, mostly the process  $\gamma^* \gamma \to \pi^+\pi^- $ has been studied
and is known to NLO precision \cite{Kivel:1999sd}. The central object
there is the two-pion distribution amplitude  
\cite{Polyakov:1999fu,Polyakov:1999gs,Polyakov:1998ze}, which has been
expressed in terms of the instanton vacuum
\cite{Polyakov:1998td}. 
The same object enters in terms
of hard diffractive electro production of two pions
\cite{Lehmann-Dronke:1999aq,Lehmann-Dronke:1999ym}. 
A  detailed QCD analyses of the cross section 
of exclusive production of pion pairs for $s< 1\;{\rm  GeV}^2$
can be found in \cite{Diehl:2000av}.
The general process $\gamma^* \gamma \to \pi^+ \pi^-$ has also been studied
earlier in the resonance region \cite{Morgan:1990kw} and in 
a purely perturbative kinematics involving two light cone wave functions
instead of a two-hadron distribution amplitude (2HDA) \cite{Brodsky:1981rp}.
\newline
\newline
In our contribution we want to look at this process from the viewpoint
of a semi-classical theory as it was proposed in the Lund model 
\cite{Andersson:1983ia}. 
In a semi-classical theory we will not be able to study the 2HDA.
However, the factorization picture motivates a semi-classical description
in which a quark-antiquark pair  produced by $\gamma^* \gamma$ 
interaction  fluctuates with some probability $P_{q\bar q \to \pi^+\pi^-}(s)$
into a $\pi^+ \pi^-$ pair. The two-meson fragmentation function 
$P_{q\bar q \to \pi^+\pi^-}(s)$ can be evaluated in terms of the string 
fragmentation picture. Decaying resonances above the mass of 
$\sqrt{s}=$1 GeV are treated here as strings. The semi classical 
picture should be a good approximation when many resonances overlap 
and interference effects can be neglected. 
\newline \newline
In the kinematics, where the above mentioned factorization holds, the common
picture is that a gluonic string is formed between the quark-antiquark pair
and finally breaks into $\pi^+ \pi^-$.
This process is of special interest because it contains  a  string
breaking exactly one tine, 
and, therefore, it would be a very interesting probe for an understanding
of the dynamics of QCD strings. 
The picture which we are referring to is an incoherent one, because we
do not work on the level of amplitudes but of probability densities.
This incoherent picture
of the Lund string model, as implemented in Monte Carlo (MC) programs 
like {\sc JETSET}  \cite{Sjostrand:1994yb}, has been very successful in
the description of many particle final states in high-energy physics. 
It is our aim
to investigate what the prediction of this model will be if we reduce
the number of final particles to very few ones, actually only two.
\newline
\newline 
In terms of the Lund model few-body states have been discussed 
only recently \cite{Andersson:1999ui}. 
The fragmentation
functions used in {\sc JETSET}  \cite{Sjostrand:1994yb} only work for high energy processes,
where the invariant mass 
of the string $\sqrt{s}$ is much larger than the masses
of the particles produced. In the two-body case the Lund model fixes the
fragmentation function only up to a normalizing function 
$g_{q\bar q}(s)= \sum_n g_{q \bar q \to n}(s)$,
where $g_{q \bar q \to n}(s)$ describes the (unnormalized)
probability that a quark-antiquark
string fragments into $n$ particles. In \cite{Andersson:1999ui} 
$g_{q\bar q}(s)$
was fixed by the requirement that for a given invariant mass squared
$s=4\; {\rm GeV}^2$ one should get  the same results as the {\sc JETSET} program, 
which originally only describes many-body states well. The crucial point is, that especially
for two particles the {\sc JETSET} program is not reliable as it
does not contain $C$-parity for example.
In this approach we try to compute $g_{q\bar q}(s)$ directly by evaluating 
the phase space and the string breaking probability for all channels
that contribute. 
\newline\newline
The paper is outlined as follows. In Sec.~\ref{sec1} we describe the factorization of
the cross section $\gamma^* \gamma \to \pi^+\pi^-$ in terms of 
Weizs\"acker-Williams
spectrum, photon structure function and the two-meson fragmentation probability.
In Sec.~\ref{sec2} we will outline how the fragmentation mechanism is understood
according to the Lund model plus some minor extensions as to spin and $C$-parity. 
Finally, in  Sec.~\ref{sec3} we will evaluate all competing 
channels to the $\pi^+\pi^-$ pair production in a region $1\;{\rm GeV}^2 < s< 2\;{\rm GeV}^2$.
We will compare our semi-classical formalism with data from the time-like pion form factor and
predict the total cross section $\gamma^* \gamma \to \pi^+\pi^-$ at LEP2 energies.
\section{The process $\gamma^* \gamma\to \pi^+ \pi^-$
 in the semi-classical theory}
\label{sec1}
\begin{figure}
\centerline{\psfig{figure=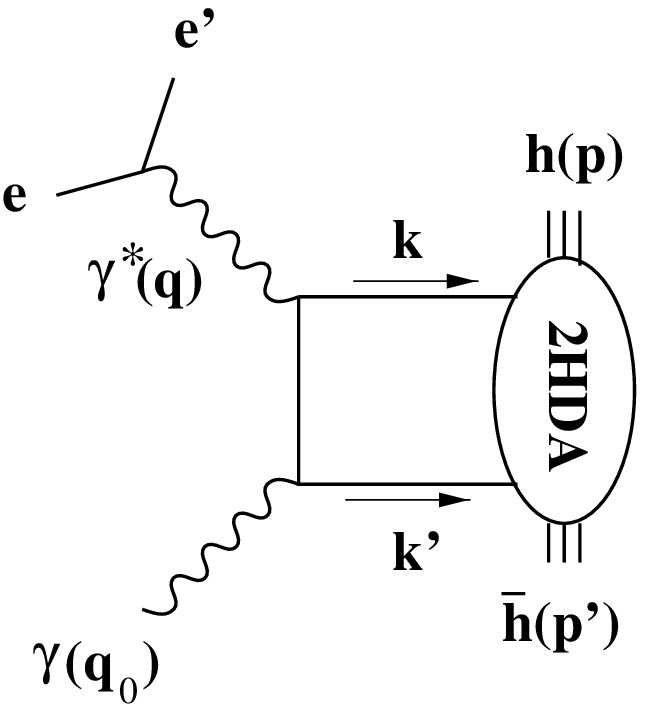,width=10cm}}
 \caption{Kinematics of the process $\gamma^* \gamma \to h\bar h$ and
definition of the two-hadron distribution amplitude (2HDA).}
\label{kinema} 
\end{figure}

\begin{figure}
\centerline{\psfig{figure=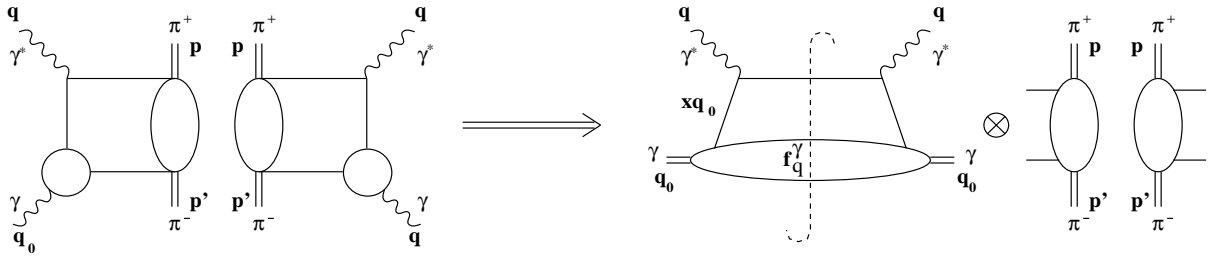,width=16cm}}
 \caption{Kinematical decomposition of the total cross section 
$\gamma^*\gamma \to \pi^+ \pi^-$ in terms of the hard scattering cross section 
$\gamma^*\gamma \to q \bar q$ and the transition probability
$q \bar q \to \pi^+ \pi^-$.}
\label{decomposition} 
\end{figure}

The kinematics of the process $\gamma^*(q) + \gamma(q_0) \to h(p)+ \bar h(p')$ 
in Fig.~\ref{kinema} has been given in 
\cite{Diehl:1998dk} and \cite{Diehl:1998pg}. The process is governed by
the virtuality of the off shell photon $Q^2 = -q^2$ and the invariant
mass squared of the final state hadrons $s = (p+p')^2$. To ensure the
factorization according to Fig.~\ref{kinema} we have to satisfy  $s \ll Q^2$.
Quark-antiquark configurations which have an invariant mass squared
$s > 1\;{\rm GeV}^2$
are treated in our approach as strings. Below this value the length of the 'string'
becomes smaller than 1 fm (with a string constant taken to be 
$\kappa \approx 1\;{\rm  GeV/fm}$), and it is not sensible to speak of the configuration 
as an extended two-dimensional object, as
it is then smaller than a typical hadron-radius. 
A more involved point is the real 
photon which enters with the momentum $q_0$. In principle it can have
a substructure in the sense that it fluctuates in to a $\rho^0$ meson 
state which interacts then with the virtual $\gamma$. This Vector Meson
Dominance (VMD) contribution is 
important  for quasi real photons  \cite{Bauer:1978iq}.
\newline\newline
In this paper
we want to look at the process $\gamma^* \gamma\to \pi^+\pi^-$
in the sense of Deep Inelastic Scattering 
(DIS) where the nucleon probe is replaced by a quasi real
photon, see e.g. \cite{Badelek:2000ea}. We treat the exclusive
process  in the spirit as the MC program {\sc LEPTO} \cite{Ingelman:1997mq}
treats the DIS of electrons off protons: The process $\gamma^* \gamma \to \pi^+ \pi^-$
factorizes in a handbag diagram including the photon structure function,
for which the whole machinery of perturbative QCD is applicable 
\cite{Witten:1977ju,Peterson:1980mz,Zerwas:1974tf,Aurenche:1996mz,Krawczyk:1997bg,Krawczyk:1998vu}, 
and the $\pi^+\pi^-$ fragmentation function, see Fig.~\ref{decomposition}.
\newline
\newline
The production of the initial $q\bar q$ pair is given by the handbag
diagram of lepton-photon scattering, while the hadronization is given
by a probability $P_{q\bar q\to \pi^+\pi^-}$ that this pair fragments into
two pions. The total cross section for the whole process 
$e^+(l_0)+ e^-(l)$
$\to$ $e^+(l_0')+ e^-(l')+ \pi^+(p) +\pi^-(p')$ 
is then given to leading order
accuracy by:  
\begin{eqnarray}
\frac{ d\sigma( e^+ e^- \to e^+e^- \pi^+ \pi^-)}{dy_0 dx dQ^2_0 dQ^2 
d^2 {\bf p_\perp}} 
&=& \sum_q e_q^2 
\frac{2\pi \alpha_{\rm em}^2}{ Q^4} (1+(1-y)^2) 
f_q^\gamma(x,Q^2,Q^2_0) 
f^T_{\gamma/e}(y_0, Q^2_0) 
P_{q \bar q \to \pi^+ \pi^-}(s,p_\perp^2)\;.
\nonumber \\
y &=& \frac{q_0 q}{q_0 l}, \quad
y_0 = \frac{q_0 l}{l_0 l}, \quad q = l-l', \quad q_0 = l_0 - l_0' \quad
Q^2 = -q^2, \quad Q_0^2 = -q_0^2, \nonumber \\
x &=& \frac{Q^2}{2 q_0q}, \quad
s = (q+q_0)^2 \approx \frac{1-x}{x}Q^2, 
\quad Q^2 = S x y y_0, \quad S= (l_0+l)^2\;.
\end{eqnarray}
%
%
%
%
$P_{q\bar q \to \pi^+\pi^-}$ depends on the transverse momentum
$p_\perp^2$ of the two produced pions,
which is defined with respect to the axis given by the string
spanned by the two initial quarks. However, $p_\perp^2$ is not a Lorentz
invariant quantity. Therefore, from an experimental point of view, 
one would like to reconstruct the
the $p_\perp^2$-dependence in terms of the Lorentz
invariant variables 
$t= (q-p)^2$ and $x$, 
see also  Eqs.~(\ref{hit1}) and (\ref{hit2}).
For the definition of the transversity it is furthermore essential, that
to leading twist we can define the quark momenta in the framework
of the standard parton model in DIS, i.e. that we can express in 
Fig.~\ref{kinema}
$k= xq_0 + q$. The virtuality $Q_0^2$ of
the photon with the 4-momentum  $q_0$ should be small, so that
it can be treated as a quasi real photon and the standard approximations 
in terms of the photo-production formalism are valid.
\newline 

Also, one should note, that in the convention  used here  
$P_{q \bar q \to \pi^+ \pi^-}(s)$ = 
$\int P_{q \bar q \to \pi^+ \pi^-}(s,p_\perp^2)d^2 {\bf p_\perp}$ 
is a dimensionless quantity.
%
%
For the factorization to be valid the ordering $Q_0^2 \ll s \ll Q^2$ must be fulfilled.
The part of the generation of the (quasi) real photon is described 
by the standard Weizs\"acker-Williams  spectrum
\cite{vonWeizsacker:1934sx,Williams:1934ad}  
for a transversely polarized photon: 
\begin{equation}
f^T_{\gamma/e}(y_0, Q^2_0) =
\frac{\alpha_{\rm em}}{2 \pi} \left( \frac{(1 + (1-y_0)^2)}{y_0}\frac{1}{Q^2_0}
-\frac{2m_e^2 y_0}{Q_0^4}\right)\;.
\end{equation}
The contribution of longitudinal polarized photons can be neglected as it
is suppressed by one power of $Q^2_0$ \cite{Friberg:1999zz}. 
\newline
\newline
The $f_q^\gamma(x,Q^2,Q^2_0)$ are the quark parton distributions of the 
quasi real photon which has the small virtuality $Q_0^2$.
%
%
We will use the parameterization given in \cite{Schuler:1995fk} 
\cite{Schuler:1996fc}. There is one subtility in using the quark parton
distributions of the photon for an approximation of the quasi real photon in our
case. Normally, those parton distributions contain also the case that the
incoming virtual photon scatters off a quark from the sea. Obviously,
in this case the 'photon remnant' is a more complicated object than just
an antiquark and it will lead in the end to some more complicated
final state than just a $\pi^+ \pi^-$ pair. As we regard exclusive
$\pi^+ \pi^-$ pair production, this part of the photon parton distribution
is not taken into account in our description. However, we recall here
the necessary condition for factorization:
\begin{equation}
x = \frac{Q^2}{Q^2 + s} \approx 1, \qquad {\rm as}\quad Q^2 \gg s \;.
\end{equation}
For large $x$, however, it is known that the parton distributions represent
nearly exclusively
the valence quarks which correspond in our case just to the valence
$q\bar q$ pair in the $\rho^0$. It means that the photon remnant is
then to a very high precision just the corresponding antiquark, 
carrying the full rest
momentum $(1-x)q_0$, and, in this respect,  the photon structure
function is a good approximation. 
\newline \newline
$P_{q \bar q \to \pi^+ \pi^-}(s)$ is
the (semi-classical) probability that the produced $q\bar q$ pair fragments
into a $\pi^+ \pi^-$ state. We will calculate this quantity in terms of the
Lund model. For intermediate values of $s$ (as compared to 
$\Lambda_{\rm QCD}^2$)
the fragmentation process should be described non-perturbatively. Here
the string picture is indisputably valid. For larger and larger $s$ a
more perturbative prescription should be valid 
\cite{Brodsky:1981rp,Diehl:1999ek}, where the second 
$q\bar q$ pair is generated
by the branching of a perturbative gluon.
In principle, the Lund model should also contain the perturbative
contribution,
because it describes the generation of the second $q\bar q$ pair in a tunneling
mechanism which should resume the interaction to all orders.
However, the model is 
an incoherent approximation, so that we cannot expect a one to one
correspondence at all. Nevertheless, it will be interesting to study how far the
string picture remains valid in this situation.
\section{String breaking and fragmentation}
\label{sec2}
In this chapter we want to derive $P_{q\bar q \to \pi^+\pi^-}$ semi-classically.
We will pick up some elements from the Lund model as described
in \cite{Andersson:1983ia}, but add extensions as to the particle  spin and $C$-parity.
The general strategy is to determine 
$P_{q\bar q \to \pi^+\pi^-}(s)$ by the fraction of the two-meson weight 
$g_{q \bar q \to \pi^+ \pi^-}(s)$  and
the total weight $g_{q\bar q}(s)$.
Here $g_{q \bar q \to \pi^+ \pi^-}(s)$ is the weight for the process that the 
initial $q \bar q$ pair fragments into a  $\pi^+ \pi^-$ pair, and $g_{q\bar q}(s)$
is the  sum of all $n$-meson  weights for all possible reactions, which the initial
$q \bar q$ pair can undergo to produce an arbitrary number of mesons.
Limiting the
value $s< 4\;{\rm GeV}^2$ we can neglect any contribution from baryons.
\begin{equation}
P_{q\bar q \to \pi^+\pi^-}(s,p_\perp^2) d^2 {\bf p_\perp} = 
\frac{g_{q \bar q \to \pi^+ \pi^-}(s,p_\perp^2) d^2 {\bf p_\perp}}
{g_{q\bar q}(s)}\;.
\end{equation}
The term 'weight' denotes a product of the phase pace of the given
process and its production probability. We will specify these objects
in the following. The simplest access is to regard the total 
weight $g_{q\bar q}(s)$. First of all, it is the sum of all 
weights of the initial $q\bar q$ pair  producing $n$ mesons:
\begin{equation}
g_{q \bar q}(s) = \sum_n g_{q \bar q \to n}(s)\;. 
\end{equation}
The $n$-particle weight is then the sum over all 
weights where the initial quark-antiquark pair $q\bar q$ fragments into
$n$ mesons $M_1,\cdots,M_n$: 
\begin{equation}
g_{q \bar q \to n}(s)  = \sum_{M_1,\cdots,M_n}  
g_{q \bar q \to M_1,\cdots,M_n}(s)\;. 
\end{equation}
In these weights only meson combinations are included which have the right
quantum number with respect to the initial state. For example, in $\gamma^* \gamma$
we have to ensure that the final hadronic state has positive $C$-parity. 
%
%
In the Lund model approach the production of final state hadrons happens
only through string breaking. This means that a resonance like $f_2(1270)$,
which is an important intermediate state for the reaction $\gamma^* \gamma \to
\pi^+\pi^-$ \cite{Lehmann-Dronke:1999aq}, is treated in the framework of
the string picture. As the string constant $\kappa$ has the value 
$\kappa \approx (1\;{\rm GeV/fm})= 0.2 \;{\rm GeV}^2$ this is a reasonable
picture for all mesons with masses above 1 GeV like the  
$f_2(1270)$ for example, because it means that the distance between the
quark-antiquark pair in the meson 
is considerable larger than 1 fm, which justifies the
picture of a string. In fact, the string picture cannot describe resonance
poles or interference effects of overlapping resonances as it is an
incoherent semi-classical picture. In fact resonances with masses
smaller than 1 GeV, namely the $\omega$ and $\rho$ resonances,
have to be treated differently as we will show
in the case of $e^+e^-\to \pi^+\pi^-$, see in Sec.~\ref{secpiform}.
Because of C-parity conservation, $\omega$ and $\rho$ resonances are
forbidden as intermediate states in exclusive $\gamma^*\gamma$ scattering.
However, they make a considerable
contribution in the reaction $\gamma^* N \to \pi^+\pi^- +N $,
see \cite{Clerbaux:2000hb}, and, as we will see later, in the 
annihilation reaction $e^+e^- \to \pi^+\pi^-$.
\newline
\newline
We start now with the description
of the two-particle weight $g_{q\bar q \to M_1 M_2}$ with $M_1$ and $M_2$
being two mesons.
\subsection{The two-particle weight}
\begin{figure}
\centerline{\psfig{figure=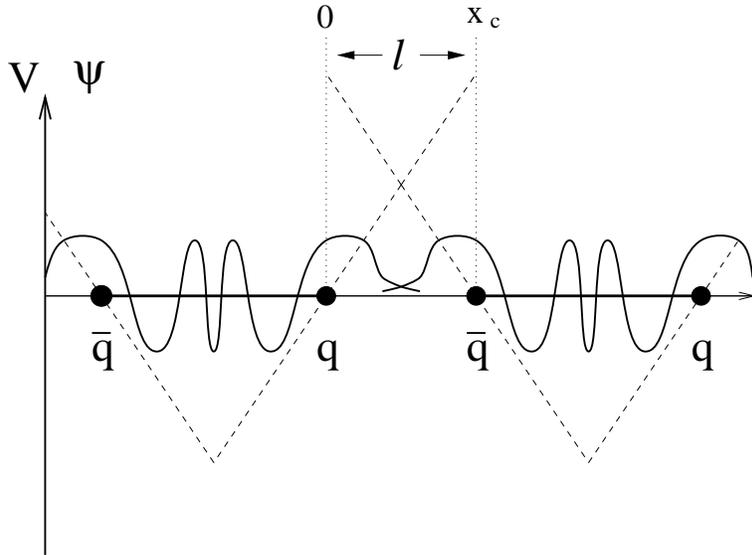,width=10cm}}
 \caption{Generation of transverse momentum $p_\perp$
through the tunneling process.
Displayed are the wave functions of the inner quark and the inner antiquark
in the linear potential of the string. The region of exponential
tunneling is of length $l = p_\perp/\kappa$, with $\kappa$ being the
string constant.}
\label{lundtunnel} 
\end{figure}

\begin{figure}
\centerline{\psfig{figure=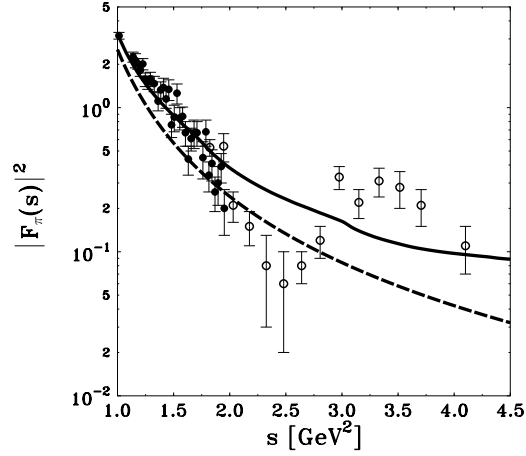,width=8.0cm}}
 \caption{Pion form factor $|F_\pi(s)|^2$ in the time-like region.
The dashed line is the contribution from the decay of a single
$\rho(770)$ resonance described in the 
VMD model. In the solid line we have added the contribution
from string fragmentation into two pions. The filled circles
are the NOVOSIBIRSK data and the open circles the data from
the DM2 collaboration.}
\label{pifofit} 
\end{figure}

\begin{figure}
\centerline{\psfig{figure=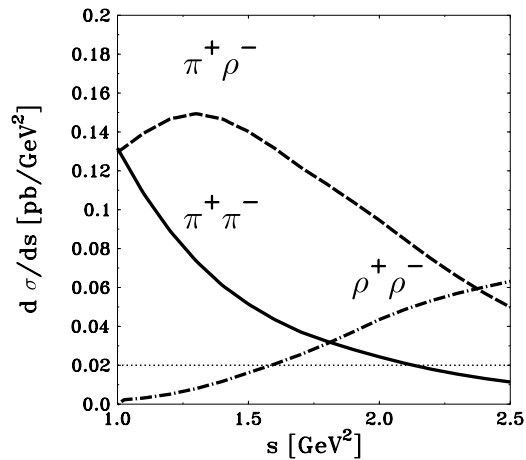,width=8.0cm}}
\caption{Cross section $d\sigma/ds(e^+e^- \to e^+ e^- A)$ 
as function of $s$ with $A=\pi^+\pi^-$
(solid line), $A = \pi^+\rho^-$ (dashed line) and $A=\rho^+\rho^-$ 
(dashed-dotted line). 
The thin dotted line is the border for 10 events per unit in
$s$ for the LEP2 luminosity (${\cal L}=500\;{\rm pb}^{-1}$). The c.m. energy
for the two electrons is $\sqrt{S} = 175 \;{\rm GeV}$.}
\label{xsec}
\end{figure}

The two-meson weight is given by the following formula:
\begin{eqnarray}
g_{q\bar q \to M_1 M_2}(s) &=& \int d^2 {\bf p_\perp} g_{q\bar q \to M_1 M_2}(s,p_\perp^2)
\nonumber \\
& =&   \int d^2 {\bf p_\perp} \int d m_1 P_{\rm mass \;M_1}(m_1) 
                       \int d m_2 P_{\rm mass \;M_2}(m_2)  
P_{\rm tunnel}(p_\perp)
\nonumber \\ && \times P_{\rm flavor}(M_1,M_2)
P_{\rm spin}(M_1)  P_{\rm spin}(M_2)
g_2(s,m_{1\perp}^2, m_{2\perp}^2)\;.
\end{eqnarray}
Here $P_{{\rm mass}\; M_1}(m)$ is the symmetric Breit-Wigner distribution to allow
for the fact that some of the mesons considered here, like the $\rho$, are
comparatively broad resonances:
\begin{equation}
P_{\rm mass}(m)  = \frac{1}{2\pi}\frac{\Gamma}{
\left[(m-m_0)^2 + \frac{\Gamma^2}{4}\right]}\;.
\end{equation}
$P_{\rm spin}(M_1)$ gives the probability that the two quarks which are under 
consideration to form the meson $M_1$ are coupled to the right spin state.
Counting the number of $S_z$ components, 
the naive values would be:
\begin{eqnarray}
P^{\rm naive}_{\rm spin}(S=0) &=& \frac{1}{4}, 
\quad 
P^{\rm naive}_{\rm spin}(S=1) = \frac{3}{4}\;.
\end{eqnarray}
However, experimentally, a ratio of pseudoscalar mesons to 
vector mesons is found that is close to 1:1 
\cite{Andersson:1983ia,Knowles:1995kj}. One can model this in assuming
that for $S=1$ only one of the three spin degrees of freedom
is active leading to the factors 
\begin{eqnarray}
P_{\rm spin}(S=0) &=& \frac{1}{4}, 
\quad 
P_{\rm spin}(S=1) = \frac{1}{4}\;.
\end{eqnarray}
$P_{\rm tunnel}$ gives the probability for the via string breaking generated
quark-antiquark pair to carry the transverse momentum $p_\perp$ with respect
to the line of the string. The transverse momentum is generated by tunneling 
through a linear barrier of length $l = p_\perp/\kappa$, see 
Fig.~\ref{lundtunnel}, with $\kappa$  being the string constant.
%
%
The tunneling process can be best understood by considering the
situation when the new quark antiquark pair has just been produced.
Then the mother string has been split into two new ones. 
Each of them consists of a quark-antiquark pair sitting in a linear potential. 
The two linear potentials
do not interfere with each other. The distance of the generated
quark-antiquark pair is $l$. The energy of the quantum mechanically forbidden
region where this quark-antiquark pair has tunneled through, has been invested
in transverse momentum $p_\perp = l\kappa$. To calculate the probability
for such a kinematical configuration one has to calculate the overlap
of the quark wave functions at the beginning and at the end of
the tunneling process.
%
%
The WKB method predicts  for the wave
function of one of the quarks in the linear potential: 
\begin{equation}
\Psi(x) =\Psi(0) \exp\left[- \int_0^x
\sqrt{ p_\perp^2 -(\kappa x')^2} dx' \right]\;.
\end{equation}
One gets the tunneling probability from the
normalized square of the overlap of the two density distributions
i.e. the square of the two wave functions:
\begin{eqnarray}
P_{\rm tunnel}( p_\perp) \sim \left| \frac{\Psi(l)}{\Psi(0)} \right|^4 
\nonumber 
\Leftrightarrow P_{\rm tunnel}( p_\perp) & = &
\frac{1}{\kappa}
\exp \left[ - \left( \frac{\pi p_\perp^2}{\kappa} \right)\right] \;.
\end{eqnarray}
For the string constant one naively would assume a value
$\kappa_{\rm string}$ $\approx$ $1\;{\rm GeV/fm}$ $\approx$ $0.2 \;{\rm GeV}^2$.
Again, it turns out in experiment that a somewhat larger value is needed
in order
to describe the $p_\perp$ of hadrons properly \cite{Andersson:1983ia}.
We get a good result in the end for the time-like pion form factor
of the pion using a value $\kappa= 0.35\;{\rm  GeV}^2$ which results
in an average hadron transverse momentum of
$\langle p_\perp \rangle = 0.472 \;{\rm GeV}$. This 
is a bit larger than the value $\langle p_\perp \rangle = 0.42 \;{\rm GeV}$
cited in \cite{Andersson:1983ia}. On the other hand, at the small
invariant masses considered here, it is likely that soft effects
which could lead to a  $p_\perp$ smearing are quite important.
\newline \newline 
$P_{\rm flavor}(M_1,M_2)$ is the probability to produce the correct flavor
for the mesons $M_1$ and $M_2$. From long termed experience with the 
{\sc JETSET} program, the following probabilities in the Lund model
have been shown to be reasonable
\cite{Andersson:1983ia,Sjostrand:1994yb}:
\begin{equation}
P_{\rm flavor}(d\bar d) = P_{\rm flavor}(u\bar u) = 1/2.3, \quad
P_{\rm flavor}(s\bar s) = 0.3/2.3 \;.
\end{equation}
The generation of heavy quarks is negligible. For high-energy fragmentation
an alternative model has been developed \cite{Chun:1998bh} using 
$P_{\rm flavor}(d\bar d)$ =$P_{\rm flavor}(u\bar u)$ =$P_{\rm flavor}(s\bar s)$
 = 1. Using this assumption, the final results for the $e^++e^-\to e^++e^-+A$
cross sections (with $A$ being a two meson system)
Fig.~\ref{xsec}  become smaller about 15\% for $\rho\rho$
and $\pi\rho$ production. In case of the production of pion pairs the
changes are negligible. The same is true for the prediction of
the time-like pion form factor Fig.~\ref{pifofit}. As the overall
changes are small we will stick in this contribution to the parameters
of the LUND model with the reservation that a precise determination of 
the strange suppression will be left to a future work on kaon production.
\newline
\newline
The central objects
of the two-particle weights are the two-particle phase space weights
$g_2(s,m_{1\perp}^2,m_{2\perp}^2)$.
$m_{j\perp}$ denotes the transverse mass, i.e., 
$m_{j\perp}^2 = m_j^2 + p_{j\perp}^2$.
For a concise treatment one has to regard the general
$n$-particle phase space weights $g_n(s,m_{1\perp}^2,\cdots,m_{n\perp}^2)$ first.
\subsection{$n$-particle weights}
\begin{figure}
\centerline{\psfig{figure=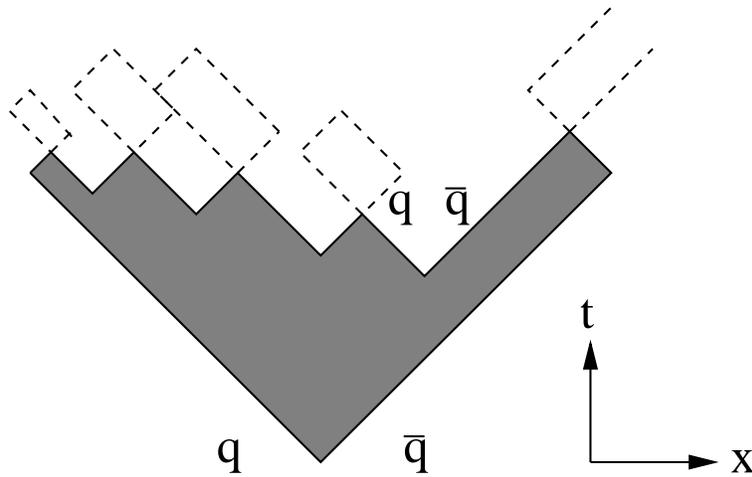,width=10cm}}
 \caption{Area law for the decay of a string into mesons in the Lund model:
The decay probability of the string is proportional to $\exp(-b{\cal A})$, with
${\cal A}$ being the shaded area  in the x-t plane shown in the figure.
The masses of the produced particles are proportional to the area
surrounded by the dashed lines.}
\label{arealaw} 
\end{figure}

\begin{figure}
\centerline{\psfig{figure=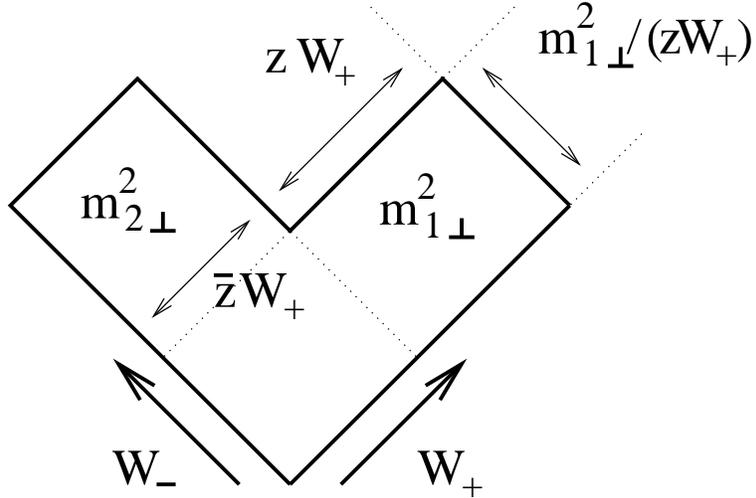,width=10cm}}
\caption{The kinematical constraints for the string breaking into two
pions. We have used the notation $\bar z = 1-z$. $W_+$ and $W_-$ are the
momenta of the initial quark and antiquark in the $t-x$ plane. $W_+W_-=s$.}
\label{yoyolc} 
\end{figure}

The $n$-particle weights $g_{q \bar q\to n}(s)$ can be derived by means of the 
$n$-particle phase space weights $g_{n}(s,m_{1\perp}^2,\dots,m_{n\perp}^2)$
which are given according to the Lund model \cite{Andersson:1983ia}:
\begin{equation}
g_{n}(s,m_{1\perp}^2,\dots,m_{n\perp}^2)
= \prod_{j=1}^n N_j d^2 {\bf p_j}
\delta(p_j^2 - m_{j\perp}^2) 
\delta \left(\sum_j p_j - (\sqrt{s},0) \right) \exp(- b {\cal A})\;.
\end{equation}
${\cal A}$ is the shaded area spanned by the particle momenta shown
in Fig.~\ref{arealaw}. $b$ is one of the two parameters in the Lund
model which have to be fitted to experimental data. 
The value used for the calculations is
$b=0.75\;{\rm GeV}^{-2}$   \cite{Bobook}.
$m_{j\perp}^2 = m_j^2 + p_{j\perp}^2$ is
the transverse mass of the j-th particle. One should observe that
the phase space weight is independent of the flavor $q$. We will
see later how $n$-particle weights and $n$-particle phase space weights
fit together. $N_j$ can be interpreted
as a sort of coupling constant for the $j$th particle. 
It can be fixed by a simple iterative condition.
The two-particle phase space weight gives
a simple analytic expression \cite{Andersson:1999ui}:
It is constructed  from energy momentum
conservation (see Fig.~\ref{yoyolc}):

\begin{equation}
m_{2\perp}^2 = (1-z)\left(s-\frac{m_{1\perp}^2}{z}\right)\quad\;,
\end{equation}
which yields:
\begin{eqnarray}
z_\pm &=& \frac{s + m_{1\perp}^2 -m_{2\perp}^2 \pm 
\sqrt{\lambda(s, m_{1\perp}^2,m_{2\perp}^2)}}{2s}
\nonumber \\
 \lambda(x,y,z) &=& x^2 + y^2 + z^2 - 2xy - 2xz - 2yz\;.
\end{eqnarray}
The area is then given by (see Fig.~\ref{yoyolc}):
\begin{equation}
{\cal A_\pm} = m_{2\perp}^2 +  \frac{m_{1\perp}^2}{z_\pm} 
=  m_{2\perp}^2 + s z_\mp 
=  \frac{1}{2}\left( s+ m_{1\perp}^2 + m_{2\perp}^2 \mp 
\sqrt{\lambda(s, m_{1\perp}^2,m_{2\perp}^2)}\right)\;.
\end{equation}
Furthermore, one has also:
\begin{equation}
\int d^2 {\bf p_1} \int d^2 {\bf p_2} \delta(p_1^2-m_{1\perp}^2) 
\delta(p_2^2-m_{2\perp}^2) \delta(p_1 + p_2 - (\sqrt{s},0))
= \frac{1}{\sqrt{\lambda(s,m_{1\perp}^2,m_{2\perp}^2)}}\;,
\end{equation}
which yields the analytic result for the two-particle phase space weight:
\begin{eqnarray}
g_{2} (s, m_{1\perp}^2,m_{2\perp}^2)
&=& N_1,N_2 \frac{2 
\exp\left[-\frac{b}{2}(s+ m_{1\perp}^2 + m_{2\perp}^2)\right]}
{b\sqrt{\lambda(s, m_{1\perp}^2,m_{2\perp}^2)}}
\cosh\left(\frac{b}{2}\sqrt{\lambda(s, m_{1\perp}^2,m_{2\perp}^2)}\right)\;.
\label{central}
\end{eqnarray}
We included a factor b in the denominator to get an dimensionless
expression.
The cosh corresponds to two possible solutions that obey energy momentum
conservation when the two-dimensional string breaks into two parts. 
Then the other phase space weights can be obtained iteratively via
\cite{Andersson:1983jt}: 
\begin{equation}
g_n (s, m_{1\perp}^2, \dots, m_{n\perp}^2)
= N_n \int\frac{dz}{z} 
\exp \left( \frac{-bm_{n\perp}^2}{z} \right)
g_{n-1} \left[(1-z)\left( s - \frac{m_{n\perp}^2}{z}\right), 
m_{1\perp}^2, \dots, m_{n-1\perp}^2\right]\;.
\end{equation}
This equation is also suitable to fix
$N_j$. In the Lund model the $N_j$ are universal constants
only dependent on $m_{j\perp}$, but not on $s$ to ensure
left right symmetry. The idea is \cite{Andersson:1983jt}
that after summation over $n$ on 
both sides, the iterative
equation has a solution in form of $\sum_n g_n(s,*) \sim s^a$  
for asymptotically large $s$, which yields then in this limit:
\begin{eqnarray}
N_j = N(a,bm_{j\perp}^2) &=& 
\lim_{s\to \infty} \left\{
1 \Bigg / \int_{m_{\perp j}^2/s}^1\frac{dz}{z} (1-z)^a
\exp \left( \frac{-bm_{j\perp}^2}{z} \right) 
\left( 1- \frac{m_{j\perp}^2}{sz}\right)^a \right\}
\nonumber \\
&=&
1 \Bigg / \int_0^1\frac{dz}{z} (1-z)^a
\exp \left( \frac{-bm_{j\perp}^2}{z} \right) 
\;.
\label{asform}
\end{eqnarray}
$a$ is the second parameter in the Lund model. The value, which
we will use in our calculation, is $a = 0.5$ \cite{Bobook}. 
\subsection{The weights for n mesons}
\begin{figure}
\centerline{\psfig{figure=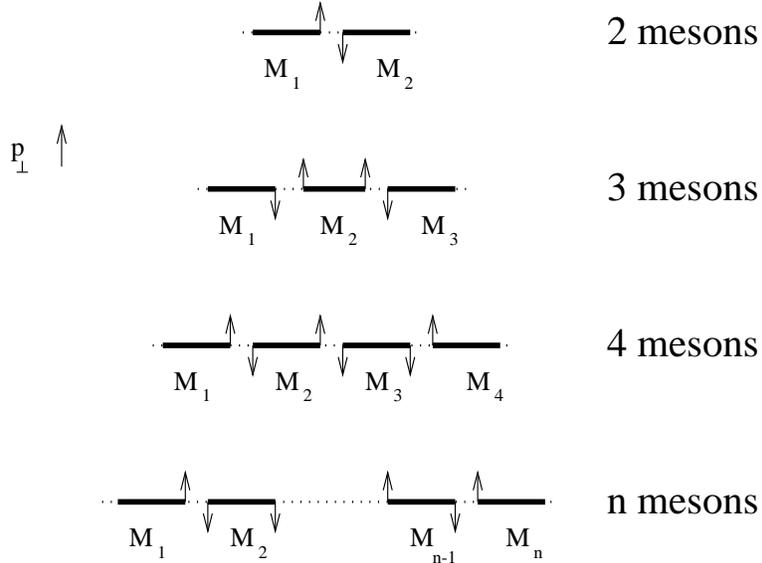,width=10cm}}
\caption{Generation of the transverse momentum via string breaking. Particles 
at the two ends of the string receive only transverse momentum from one breakpoint, 
while those in the middle receive transverse momentum from two. So, the average 
transverse momentum for particles at the end is $\kappa/\pi$ while
it is $2\kappa/\pi$ for particles in a mid position.}
\label{string3}
\end{figure}

\begin{figure}
\centerline{\psfig{figure=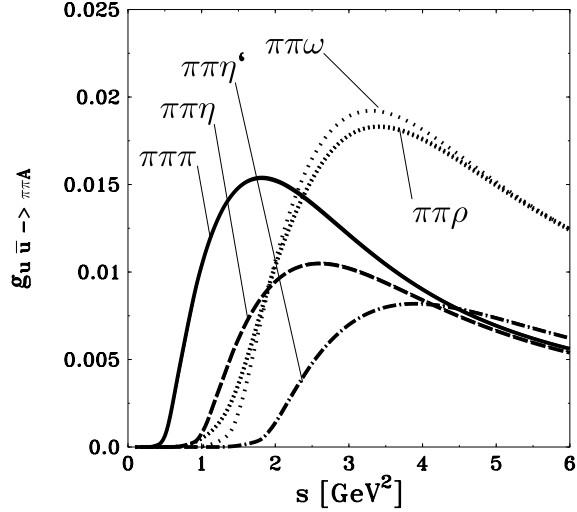,width=9cm}}
 \caption{The three-particle weights
 $g_{q\bar q \to \pi \pi A}$ as functions of $s$, 
The following
contributions are displayed:\\
$g_{u\bar u/ d\bar d \to \pi^0\pi^0\pi^0}(s)$,
$g_{u\bar u/ d\bar d \to \pi^+\pi^-\pi^0}(s)$ (solid); \\
$g_{u\bar u/ d\bar d \to \pi^0\pi^0\eta}(s)$,
$g_{u\bar u/ d\bar d \to \pi^+\pi^-\eta}(s)$ (dashed); \\
$g_{u\bar u/ d\bar d \to \pi^0\pi^0\eta'}(s)$,
$g_{u\bar u/ d\bar d \to \pi^+\pi^-\eta'}(s)$ (dashed-dotted); \\
$g_{u\bar u/ d\bar d \to \pi^+\pi^-\rho^0}(s)$,
$g_{u\bar u/ d\bar d \to \pi^+\pi^-\omega}(s)$,
$g_{u\bar u/ d\bar d \to \pi^-\pi^0\rho^+}(s)$,
$g_{u\bar u/ d\bar d \to \pi^+\pi^0\rho^-}(s)$
 (bold dotted).}
\label{g3pipiX} 
\end{figure}

The procedure described in the previous two subsections can 
easily be generalized to describe the weight to produce $n$ mesons
$M_1,\cdots, M_n$ from one mother string:
\begin{eqnarray}
g_{q \bar q \to M_1,\dots M_n}(s) &=&  
\left[\prod_{j=1}^{n-1} \int d^2 {\bf p_{j\perp}}
P_{\rm tunnel}(p_{j\perp})\right]
\left[\prod_{i=1}^{n}\int d m_i P_{\rm mass, M_i}(m_i)
P_{\rm spin}(M_i) \right]
\nonumber \\ &\times&
 g_n(s,m_{1\perp}^2,\dots,m_{n\perp}^2) P_{\rm flavor}(M_1,\dots,M_n)\;.
\end{eqnarray}
One should then observe that:
\begin{eqnarray}
m_{1\perp}^2 &=& m_1^2 + p_{1\perp}^2\; ;
\nonumber \\
m_{i\perp}^2 &=& m_i^2 + (p_{i-1\perp}- p_{i\perp})^2, \quad (1<i<n)\; ;
\nonumber \\
m_{n\perp}^2 &=& m_n^2 + p_{n-1\perp}^2\; .
\end{eqnarray}
In practice, it will be easier to make use of the recurrence relation. Here
the three-meson phase space weight is given by:
\begin{equation}
g_3(s,m_{1\perp}^2,m_{2\perp}^2,m_{3\perp}^2) 
= N(a,bm_{3\perp}^2)
\int\frac{dz}{z} \exp\left(-\frac{bm_{3\perp}^2}{z} \right)
g_2\left((1-z)\left[s-\frac{m_{3\perp}^2}{z}\right],m_{1\perp}^2,m_{2\perp}^2\right)\;.
\end{equation} 
What causes problems numerically, 
is the integration over the transverse momenta.
The kinematic situation is the one of Fig.~\ref{string3}, i.e., 
that the particle in 
the middle receives transverse momentum from
two string break points, whereas a particle at the end only from one. 
To simplify things we will take average values instead of a time 
consuming $p_\perp$ integration. First, the average over the 
relative orientation of the transverse vectors yields: 
\begin{equation}
(p_{1\perp} - p_{2\perp})^2 \approx p_{1\perp}^2 +   p_{2\perp}^2\;.
\end{equation}
To simplify things further, we will also replace 
$p_{1\perp}^2\approx \langle p_{1\perp}^2\rangle = \kappa/\pi$ for
one of the three particles. 
Taking into account that each of the  three particles may sit in mid 
position in Fig.~\ref{string3}, we get:
\begin{eqnarray}
&&
g_{q\bar q \to M_1M_2M_3}(s,m_1^2,m_2^2,m_3^2)\
\nonumber \\ &=& 
\int\frac{dz}{z} \Bigg[ 
e^{-\frac{b(m_3^2 + 2\kappa/\pi)}{z}} N(a,b(m_3^2+2\kappa/\pi)) 
g_{q\bar q \to M_1M_2}\left((1-z)\left(s-\frac{m_3^2 + 2\kappa/\pi}{z}\right),m_1^2,m_2^2\right)
\nonumber \\
&&\qquad + 
e^{-\frac{b(m_3^2 + \kappa/\pi)}{z}} N(a,b(m_3^2+\kappa/\pi))
g_{q\bar q \to M_1M_2}\left((1-z)\left(s-\frac{m_3^2 + \kappa/\pi}{z}\right),m_1^2+\kappa/\pi,m_2^2\right)
\nonumber \\
&& \qquad + 
e^{-\frac{b(m_3^2 + \kappa/\pi)}{z}} N(a,b(m_3^2+\kappa/\pi))
g_{q\bar q \to M_1M_2}\left((1-z)\left(s-\frac{m_3^2 + \kappa/\pi}{z}\right),m_1^2,m_2^2+\kappa/\pi\right)
\Bigg]
\nonumber \\
&& \times P_{\rm spin}(M_3) P_{\rm flavor}\;.
\label{threeparticle}
\end{eqnarray}
Here, we have taken advantage of the fact that
the functions $g_n(s,m^2_{1\perp},\dots,m^2_{n\perp})$
are symmetric under permutation of the $m_{i,\perp}^2$ up to 
a finite energy correction. We will check later to which extent this
symmetry is fulfilled. Furthermore, we have to state that
this approximation is only valid  because we will restrict
ourselves to $u$ and $d$-quarks. Strange quarks would lead
to at least two kaons plus a third meson and this
will not contribute in the range $1\;{\rm GeV}^2<s<2\;{\rm GeV}^2$.
%
%
%
One has to take into account that one has to deal with transverse
masses here, i.e. one has to add the average transverse momentum, which
is roughly 1/3 GeV per contributing quark in our case. If one calculates
the weight for KK$\pi$ production, for example, one will observe that its
shape resembles essentially the one of the $\pi\pi\eta'$ contribution
in Fig.~\ref{g3pipiX}, which starts around $s=2\;{\rm GeV}^2$.
Neglecting the KK$\pi$ and corresponding channels,
the expression Eq.~(\ref{threeparticle}) is basically flavor independent, as u and d quarks
are treated equal because of their small mass difference.
%
%
Otherwise we would be in trouble because 
the initial quark-antiquark pair in the two-meson weight and 
the three-meson weight are not identical. Our three-meson 
weight contains then all three possible combinations of the 
three particles being formed along the line of a string breaking two times. 
One finds analogously for the four-meson weight:
\begin{eqnarray}
&&
g_{q\bar q \to M_1M_2M_3M_4}(s,m_1^2,m_2^2,m_3^2,m_4^2)\
\nonumber \\ &=& 
\int\frac{dz}{z} \Bigg[ 
2e^{-\frac{b(m_4^2 + 2\kappa/\pi)}{z}} N(a,b(m_4^2+2\kappa/\pi))
g_{q\bar q \to M_1M_2M_3}\left((1-z)\left(s-\frac{m_4^2 + 2\kappa/\pi}{z}\right),m_1^2,m_2^2,m_3^2\right)
\nonumber \\
&&\qquad + 
\frac{2}{3} e^{-\frac{b(m_4^2 + \kappa/\pi)}{z}} N(a,b(m_4^2+\kappa/\pi))
g_{q\bar q \to M_1M_2M_3}\left((1-z)\left(s-\frac{m_4^2 + \kappa/\pi}{z}\right),m_1^2+\kappa/\pi,m_2^2,m_3^2\right)
\nonumber \\
&& \qquad + 
\frac{2}{3}e^{-\frac{b(m_4^2 + \kappa/\pi)}{z}} N(a,b(m_4^2+\kappa/\pi))
g_{q\bar q \to M_1M_2M_3}\left((1-z)\left(s-\frac{m_4^2 + \kappa/\pi}{z}\right),m_1^2,m_2^2+\kappa/\pi,m_3^2\right)
\nonumber \\
&& \qquad + 
\frac{2}{3}e^{-\frac{b(m_4^2 + \kappa/\pi)}{z}} N(a(b,m_4^2+\kappa/\pi))
g_{q\bar q \to M_1M_2M_3}\left((1-z)\left(s-\frac{m_4^2 + \kappa/\pi}{z}\right),m_1^2,m_2^2,m_3^2+\kappa/\pi\right)
\Bigg]
\nonumber \\
&& \times P_{\rm spin}(M_4) P_{\rm flavor}\;.
\label{fourparticle}
\end{eqnarray}
The factors take into account that the fourth particle has two possibilities 
to sit at the head of the string
and two possibilities to sit the midst, see Fig.~\ref{string3}.

\section{Numerical estimates}
\label{sec3}
\subsection{Two-meson weights}
\begin{figure}
\centerline{\psfig{figure=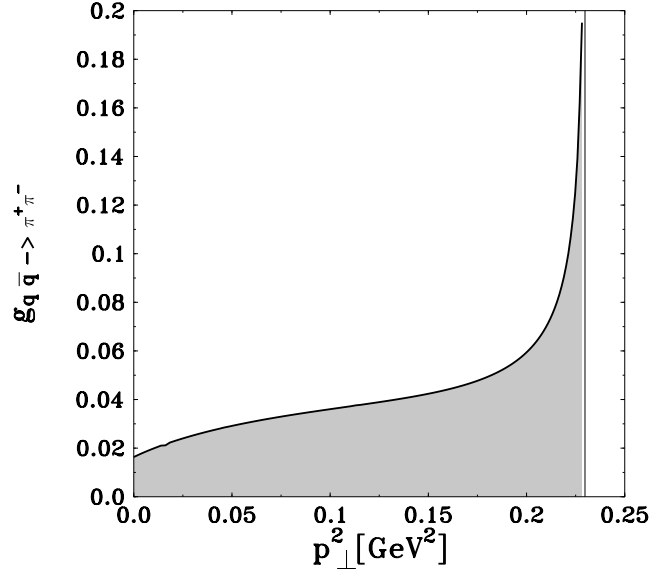,width=10cm}}
 \caption{The weight $g_{u\bar u \to \pi^+\pi^-}
(s=1\;{\rm GeV}^2,p_\perp^2)$ as function of $p_\perp^2$. 
The shaded area indicates the allowed region
$s> 4 m_{\pi\perp}^2$ on the $x$-axis. 
The singularity at $s=4 m_{\pi\perp}^2$ is integrable.} 
\label{gupipipt} 
\end{figure}

\begin{figure}
\centerline{\psfig{figure=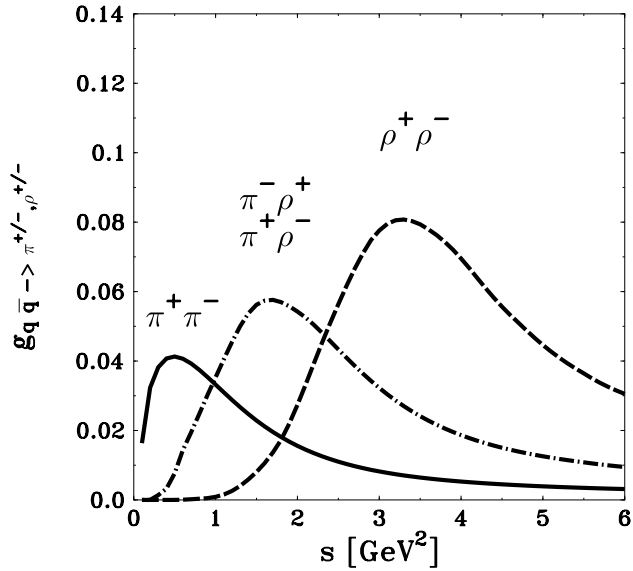,width=10cm}}
 \caption{The weights $g_{q\bar q \to \pi^\pm,\; \rho^\pm}$
as functions of $s$. The 
solid line shows the weights
$g_{u\bar u \to \pi^+\pi^-}$,
$g_{d\bar d \to \pi^+\pi^-}$, and
the dashed line  
$g_{u\bar u \to \rho^+\rho^-}$,
$g_{d\bar d \to \rho^+\rho^-}$.
The dashed-dotted line represents the four weights
$g_{u\bar u \to \pi^+\rho^-}$, $g_{u\bar u \to \rho^+\pi^-}$,
$g_{d\bar d \to \pi^+\rho^-}$, $g_{d\bar d \to \rho^+\pi^-}$.}
\label{chns} 
\end{figure}

\begin{figure}
\centerline{\psfig{figure=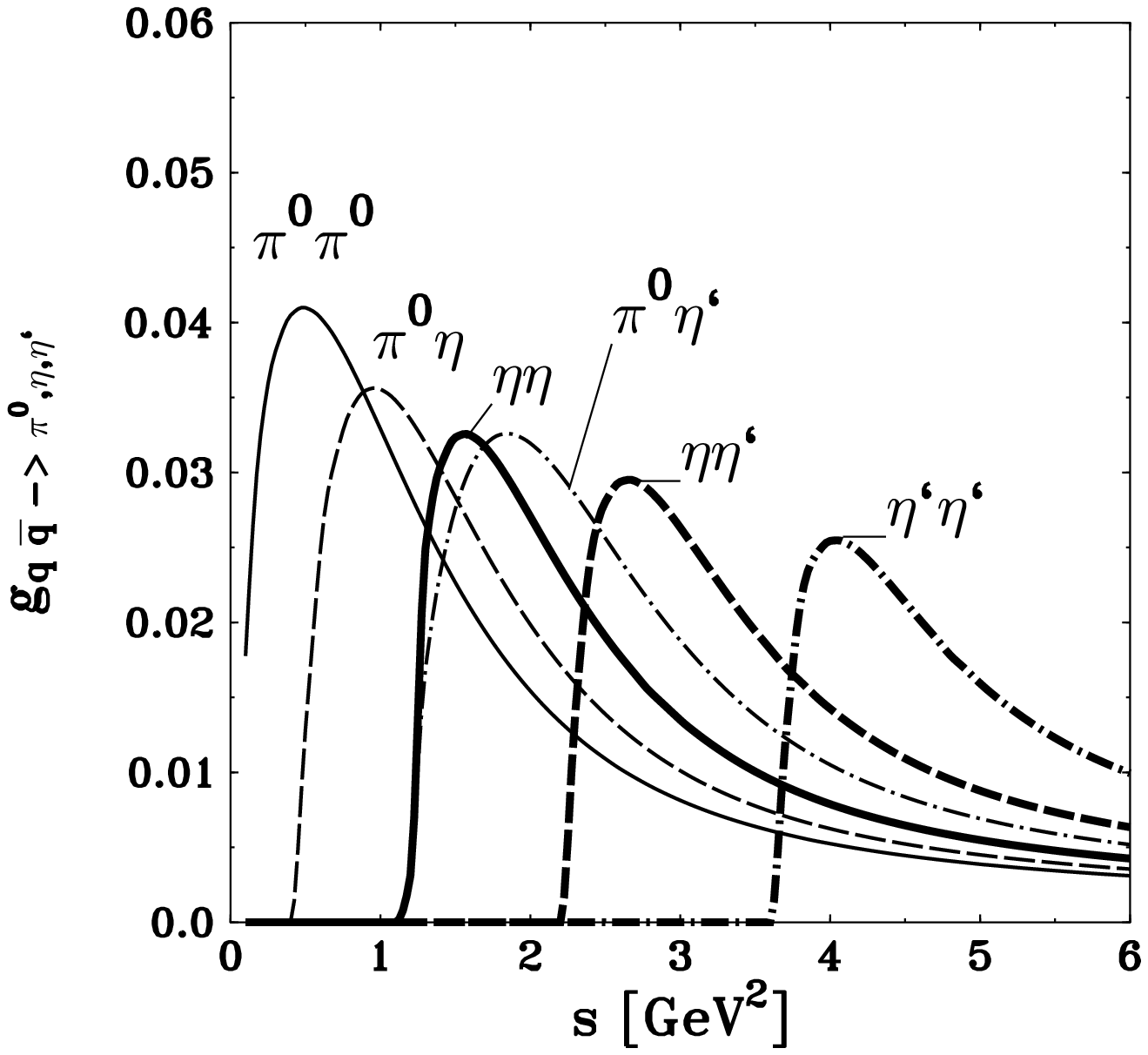,width=9cm}}
 \caption{The weights $g_{q\bar q \to \pi^0,\eta,\eta'}$
as functions of $s$, i.e. the
contribution of neutral ($C=+$) pseudoscalar mesons.
The following
contributions are displayed:\\
$g_{u\bar u \to \pi^0\pi^0}$, $g_{d\bar d\to \pi^0\pi^0}$ (thin solid); \\
$g_{u\bar u \to \pi^0\eta}$, $g_{d\bar d\to \pi^0\eta}$ (thin dashed); \\
$g_{u\bar u \to \pi^0\eta'}$, $g_{d\bar d\to \pi^0\eta'}$ (thin dashed-dotted); \\
$g_{u\bar u \to \eta\eta}$, $g_{d\bar d\to \eta\eta}$ (bold solid); \\
$g_{u\bar u \to \eta\eta'}$, $g_{d\bar d\to \eta\eta'}$ (bold dashed); \\
$g_{u\bar u \to \eta'\eta'}$, $g_{d\bar d\to \eta'\eta'}$ (bold dashed-dotted).}
\label{ncnscplus} 
\end{figure}

\begin{figure}
\centerline{\psfig{figure=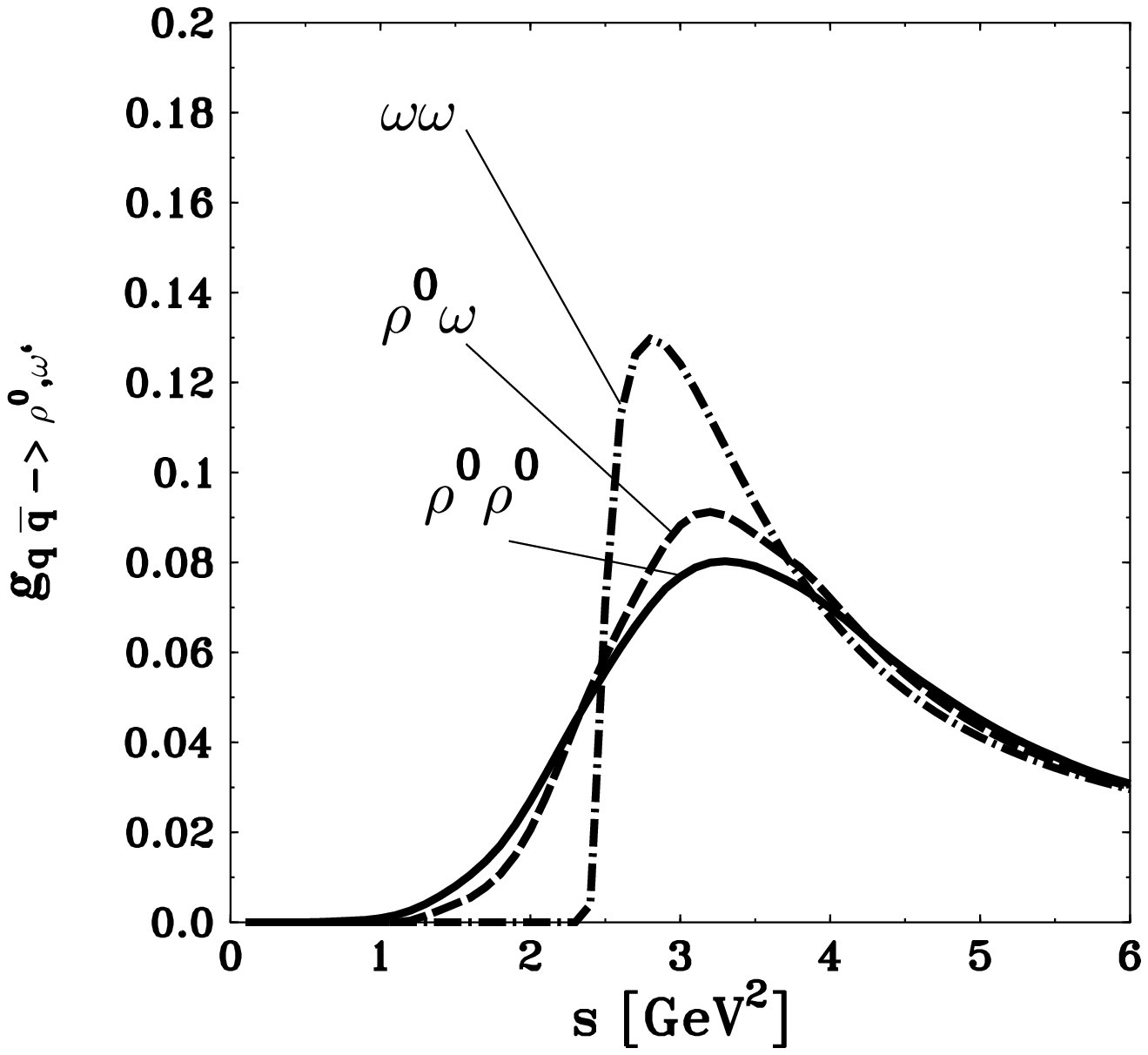,width=9cm}}
 \caption{The weights $g_{q\bar q \to \rho^0,\;\omega}$
as functions of $s$, i.e. the
contribution of neutral ($C=-$) vector mesons.
The following
contributions as functions of $s$ are displayed:\\
$g_{u\bar u \to \rho^0\rho^0}$, $g_{d\bar d\to \rho^0\rho^0}$ (solid); \\
$g_{u\bar u \to \rho^0\omega}$, $g_{d\bar d\to \rho^0\omega}$ (dashed); \\
$g_{u\bar u \to \omega\omega}$, $g_{d\bar d\to \omega\omega}$ (dashed-dotted).}
\label{ncnscminus} 
\end{figure}

\begin{figure}
\centerline{\psfig{figure=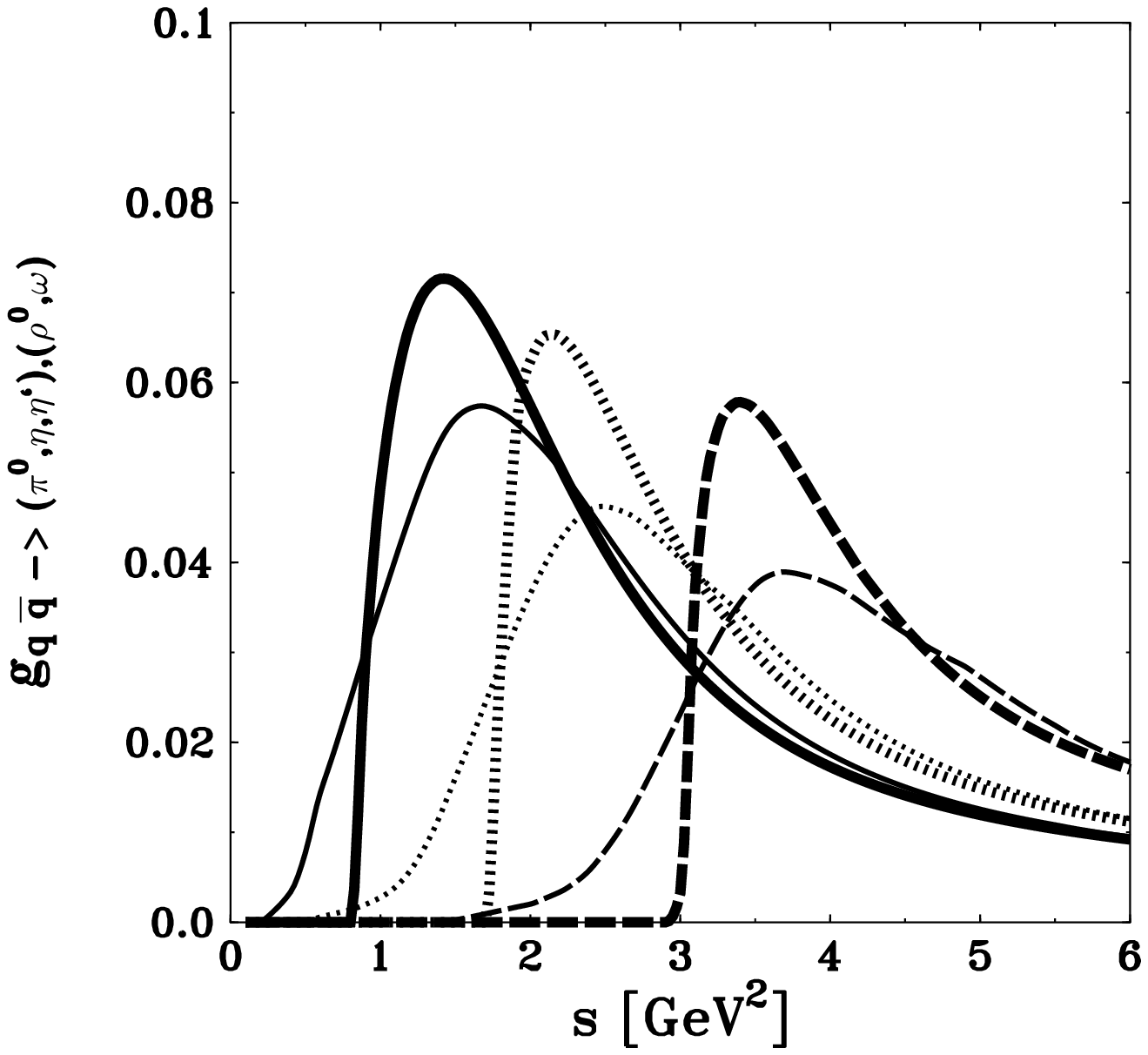,width=8.5cm}}
 \caption{The weights $g_{q\bar q \to (\pi^0,\eta ,\eta'),(\rho^0,\omega)}$
as functions of $s$, i.e. the
contribution of a pair of one pseudoscalar and one vector meson. 
The following
contributions are displayed:\\
$g_{u\bar u \to \pi^0\rho^0}$, $g_{d\bar d\to \pi^0\rho^0}$ (thin solid); \\
$g_{u\bar u \to \pi^0\omega}$, $g_{d\bar d\to \pi^0\omega}$ (bold solid); \\
$g_{u\bar u \to \eta\rho^0}$, $g_{d\bar d\to \eta\rho^0}$ (thin dotted); \\
$g_{u\bar u \to \eta\omega}$, $g_{d\bar d\to \eta\omega}$ (bold dotted); \\
$g_{u\bar u \to \eta'\rho^0}$, $g_{d\bar d\to \eta'\rho^0}$ (thin dashed); \\
$g_{u\bar u \to \eta'\omega}$, $g_{d\bar d\to \eta'\omega}$ (bold dashed).}\
\label{ncnscmixed} 
\end{figure}

\begin{figure}
\centerline{\psfig{figure=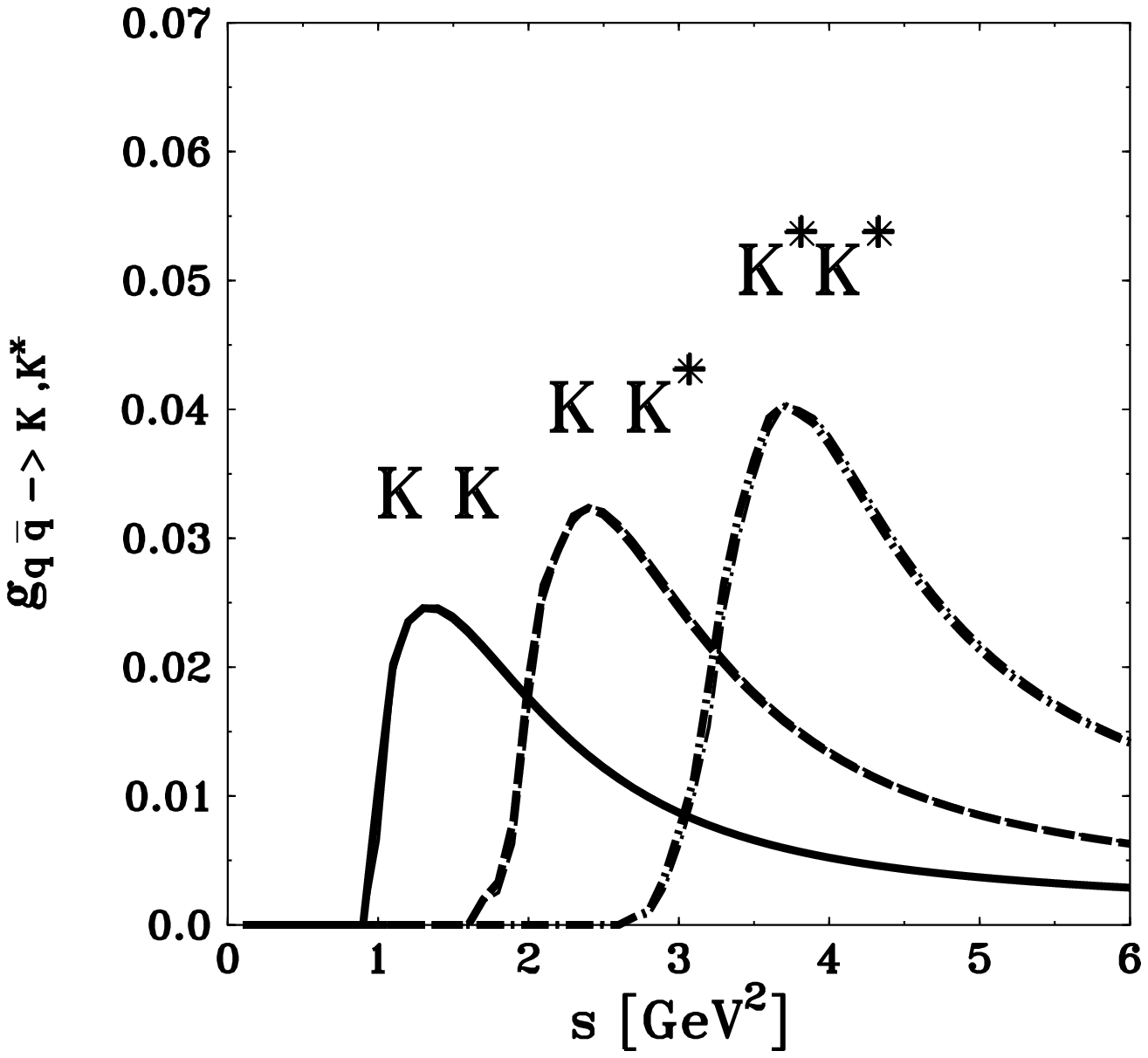,width=8.5cm}}
 \caption{The weights $g_{q\bar q \to K,K^{*}}$
as functions of $s$, i.e. the
contribution of neutral and charged kaons. Because of being
close together in their mass it is not possible to distinguish
in the plot between charged and neutral kaons.
The following
contributions are displayed:\\
$g_{d\bar d \to K^0 K^0}$, $g_{u\bar u \to K^+ K^-}$ (solid); \\
$g_{d\bar d \to K^0 K^{*0}}$, $g_{u\bar u \to K^+ K^{*-}}$, 
$g_{u\bar u \to K^- K^{*+}}$ (dashed); \\
$g_{d\bar d \to K^{*0} K^{*0}}$, $g_{u\bar u \to K^{*+} K^{*-}}$ 
(dashed-dotted).}
\label{st} 
\end{figure}

\begin{figure}
\centerline{\psfig{figure=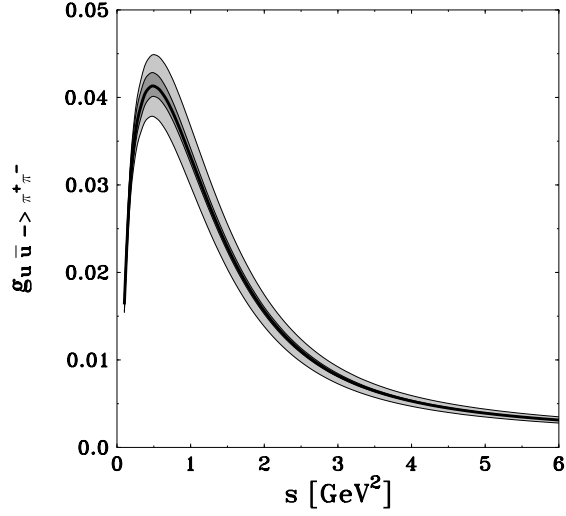,width=9cm}}
 \caption{Dependency of the function $g_{u\bar u \to \pi^+\pi^-}(s)$
on the parameters $a$ and $b$. The bold solid line shows the result
for $a=0.5$ and $b=0.75\;{\rm GeV}^{-2}$ as used in all calculations.
The light grey area shows the variation  from $a=0.4$ to $a=0.6$, while
the dark grey area shows the variation from  $b=0.65\;{\rm GeV}^{-2}$ to
$b=0.85\;{\rm GeV}^{-2}$.}
\label{checkparam} 
\end{figure}

Fig.~\ref{gupipipt} shows the weight 
$g_{u\bar u \to \pi^+\pi^-}(s= 1\;{\rm  GeV}^2,p_\perp^2)$. The region
in $p_\perp^2$ is limited by the requirement that 
$\lambda^2(s,m_{1\perp}^2,m_{2\perp}^2)>0$. One encounters from the 
phase-space an integrable singularity at:
\begin{equation}
p_\perp^2 = \frac{1}{4}\left[ s - 2 (m_1^2+m_2^2) 
+ \frac{(m_1^2-m_2^2)^2}{s}\right]\;.
\end{equation}
We have to calculate all possible two-meson weights. As we will restrict
to a comparatively small region in s, i.e. $1\;{\rm GeV}^2<s<2\;{\rm GeV}^2$,
we will only take into account the lightest pseudoscalar mesons 
and vector mesons. As we are only interested in $\pi^+\pi^-$ production we
have to regard for the principal quark-antiquark pair only the flavors 
$u$ and $d$. As heavy quarks are actually not generated in fragmentation
we can exclude all mesons carrying charm, bottom and of course top quarks.
In our formalism, an $s\bar s$ pair can only be generated through string
breaking, and consequently, it cannot recombine  to a $\Phi$ meson,
as the two strange quarks belong to different strings.
Therefore, the generation of
an $s\bar s$ pair in fragmentation leads exclusively to kaon pairs, while
the $d\bar d$ and $u\bar u$ pairs only produce pions, rhos, etas, and omegas.
$\Phi$ mesons are neglected altogether because they consist  to almost 100\% of 
$s\bar s$ \cite{Caso:1998tx}. According to recent analysis 
\cite{Bramon:1999va,Feldmann:1998vh,Feldmann:2000uf},
the $\eta$ meson consists to 40\% of $s\bar s$ and the $\eta'$ meson
to 60\% of $s\bar s$. This means that we have to weight the $\eta$
production by a factor 3/5 and the $\eta'$ production by a factor 2/5.
\newline
\newline
For neutral particles that are their own antiparticles
we have to take into account that the final state must be in total $C=+$.
This leads to the possible combinations in Tab.~\ref{allowedpairs}.
The masses and widths used for the computation are given in 
Tab.~\ref{widthmass}.
In fact, the Breit-Wigner distribution has to be taken into account only for
the vector mesons.
\newline
\newline
Fig.~\ref{chns} shows the contributions from the charged non strange
sector, i.e. the combinations $\pi^\pm,\rho^\pm$. 
In general, vector mesons have a larger mass than their
pseudoscalar partners. Therefore, 
their weight starts later in $s$,
where there is more competition with other channels. One sees later that 
this suppresses the contribution of the $\rho$. 
\newline
\newline
Fig.~\ref{ncnscplus} shows the two-meson weights for the neutral 
$C=+$ mesons. One observes that due to the strange content of the
$\eta$ and $\eta'$ mesons their weight is suppressed.
The neutral $C=-$ sector consists only of combinations of $\rho^0$ and
$\omega$ mesons. Their masses are close to each other, so the only
effect visible comes from their different widths, see Fig.~\ref{ncnscminus}.
In case of  the mixed combinations 
(one vector meson and one pseudoscalar meson) we observe both effects: the
suppression of the contribution from the $\eta$ and $\eta'$ mesons on
the one hand and the fact, that $\rho^0$ and $\omega$ mesons basically
can be only distinguished according to their widths, on the other hand
(Fig.~\ref{ncnscmixed}). The mixed combinations
 will not contribute in $\gamma^* \gamma$ scattering
because of $C$-parity conservation, but they will contribute to other
processes like $e^+e^-$ annihilation. 
For the strange meson contribution
(Fig.~\ref{st}) one observes the suppression of the strangeness production
versus the production of $u\bar u$ and $d\bar d$ pairs.
\newline\newline
It is very interesting to investigate the dependence of our results
on the parameters $a$ and $b$ in the Lund model. In Fig.~\ref{checkparam}
we show again the function $g_{u\bar u \to \pi^+\pi^-}(s)$ with $a$ varying
between $0.4<a<0.6$ and $b$ varying between 
$0.65\;{\rm  GeV}^{-2}<b<0.85\;{\rm  GeV}^{-2}$. The
parameter $a$ accounts for most of the uncertainty.
In general, the deviations are small. This  means that the
results are relatively insensitive to the choice of $a$ and $b$. 
\newline
\newline
From the results obtained so far,
we can compute the total two-meson  or two-particle
weight function. In case of a $C=+$ state it is given by:
\begin{eqnarray}
g^{(+)}_{u\bar u \to 2}(s) &=& g_{u\bar u \to \pi^+\pi^-}(s)
                        +g_{u\bar u \to \pi^+\rho^-}(s)
                        +g_{u\bar u \to \rho^+\pi^-}(s)
                        +g_{u\bar u \to \rho^+\rho^-}(s)
\nonumber \\
&&    +g_{u\bar u \to \pi^0\pi^0}(s)
      +g_{u\bar u \to \eta\eta}(s)             
      +g_{u\bar u \to \eta'\eta'}(s)             
\nonumber \\
&&  +2g_{u\bar u \to \pi^0\eta}(s)
    +2g_{u\bar u \to \pi^0\eta'}(s)             
    +2g_{u\bar u \to \eta\eta'}(s)      
\nonumber \\
&&    +g_{u\bar u \to \rho^0\rho^0}(s)
      +g_{u\bar u \to \omega\omega}(s)             
      +2g_{u\bar u \to \rho^0\omega}(s)
\nonumber \\
&& + g_{u\bar u \to K^+K^-}(s)
   + g_{u\bar u \to K^+K^{*-}}(s)
   + g_{u\bar u \to K^{*+}K^-}(s)
   + g_{u\bar u \to K^{*+}K^{*-}}(s)\;;
\end{eqnarray}
and in case of a $C=-$ state by:
\begin{eqnarray}
g^{(-)}_{u\bar u \to 2}(s) &=& g_{u\bar u \to \pi^+\pi^-}(s)
                              +g_{u\bar u \to \pi^+\rho^-}(s)
                              +g_{u\bar u \to \rho^+\pi^-}(s)
                              +g_{u\bar u \to \rho^+\rho^-}(s)
\nonumber \\
&&    +2g_{u\bar u \to \pi^0\rho^0}(s)
      +2g_{u\bar u \to \eta\rho^0}(s)             
      +2g_{u\bar u \to \eta'\rho^0}(s)             
\nonumber \\
&&    +2g_{u\bar u \to \pi^0\omega}(s)
      +2g_{u\bar u \to \eta\omega}(s)             
      +2g_{u\bar u \to \eta'\omega}(s)             
\nonumber \\
&& + g_{u\bar u \to K^+K^-}(s)
   + g_{u\bar u \to K^+K^{*-}}(s)
   + g_{u\bar u \to K^{*+}K^-}(s)
   + g_{u\bar u \to K^{*+}K^{*-}}(s)\;.
\end{eqnarray}
In the case $g^{(\pm)}_{d\bar d \to 2}(s)$ only the contribution of the
charged kaons has to be replaced by neutral kaons. 
\subsection{Three-meson weights}
\begin{figure}
\centerline{\psfig{figure=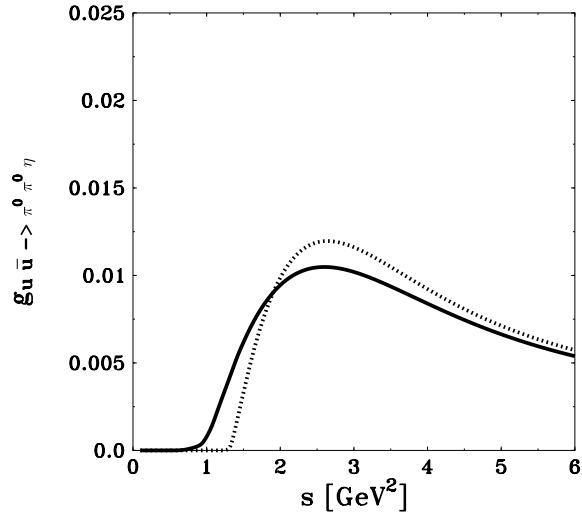,width=9cm}}
 \caption{Error of the approximation for the three-meson
weights. The plot shows the weight 
$g_{u\bar u\to \pi^0\pi^0\eta}(s)$
calculated in two different ways.
The solid line shows the form used in the calculations, where the heavier
 $\eta$ meson
is approximated. The dotted line shows the same result, but in a case
where one of the two $\pi^0$ has been approximated.}
\label{symmetry} 
\end{figure}

For the three-meson weights we make use of the approximation 
Eq.~(\ref{threeparticle}). 
We will only take into consideration the lightest
mesons. In case of neutral particles we get again limitations because of 
$C$-parity conservation, thus reducing the number of possible
combinations to the following for a $C=+$ state:
\begin{center}
\begin{tabular}{lll}
$\pi^+\pi^- \pi^0,$&$\pi^0\pi^0\pi^0,$ &\\
$\pi^+\pi^- \eta, $&$ \pi^0\pi^0\eta, $ &\\
$\pi^+\pi^- \eta', $&$ \pi^0\pi^0\eta', $& \\
$\pi^+\pi^- \rho^0, $&$ \pi^+\pi^-\omega,$&\\
$\pi^+\pi^0\rho^-, $ &$ \pi^-\pi^0\rho^+;$ & \\
\end{tabular}
\end{center}
and to 
\begin{center}
\begin{tabular}{lll}
$\pi^+\pi^- \rho^0,$&$\pi^0\pi^0\rho^0,$ &\\
$\pi^+\pi^- \omega, $&$ \pi^0\pi^0\omega, $ &\\
$\pi^+\pi^- \pi^0, $& \\
$\pi^+\pi^- \eta, $&$ \pi^+\pi^-\eta', $& \\
$\pi^+\pi^0\rho^-, $ &$ \pi^-\pi^0\rho^+$ & \\
\end{tabular}
\end{center}
for a $C=-$ state.
We put the heaviest particle always at position 3 in 
Eq.~(\ref{threeparticle}) because this is the position where the 
approximation $p_\perp^2 \approx \langle p_\perp^2\rangle = \kappa/\pi$
takes place and the error should be proportional to $p_\perp^2/m^2$.
Fig.~\ref{g3pipiX} shows the various three-meson contributions. 
A quite substantial contribution comes from the combinations $\pi\pi \rho$
and $\pi\pi \omega$.
\newline
\newline
It is important to make a check of the reliability of the approximation.
In Eq.~(\ref{threeparticle}) the meson in the position 3 is treated different
from the mesons at position 2 and 1. We will have a brief look what consequences
this asymmetry has. In Fig.~\ref{symmetry} we show for the combination 
$\pi^0\pi^0\eta$ the two possibilities: One time the $\eta$ holds position
three as in the calculation and one time it is on position 2 or 1, respectively.
One observes that the absolute magnitude does not change much, but
that when $\eta$ sits on position 2 or 1 the weight starts later in $s$. 
From the calculated three-meson weights we can compute the three-particle
weight, which is in case of a $C=+$ state:
\begin{eqnarray}
g^{(+)}_{u\bar u \to 3}(s) &=&  g_{u\bar u \to \pi^+\pi^-\pi^0}(s)
                         +g_{u\bar u \to \pi^0\pi^0\pi^0}(s)
\nonumber \\
                   &&    +g_{u\bar u \to \pi^+\pi^-\eta}(s)
                         +g_{u\bar u \to \pi^0\pi^0\eta}(s)
\nonumber \\
                   &&    +g_{u\bar u \to \pi^+\pi^-\eta'}(s)
                         +g_{u\bar u \to \pi^0\pi^0\eta'}(s)
\nonumber \\
                   &&    +g_{u\bar u \to \pi^-\pi^0\rho^+}(s)
                         +g_{u\bar u \to \pi^+\pi^0\rho^-}(s)
                         +g_{u\bar u \to \pi^+\pi^-\rho^0}(s)
                         +g_{u\bar u \to \pi^+\pi^-\omega}(s)\;.
\end{eqnarray}                                                                  
The corresponding expression for the $C=-$ state is trivial.
\subsection{Four-meson weights}
\begin{figure}
\centerline{\psfig{figure=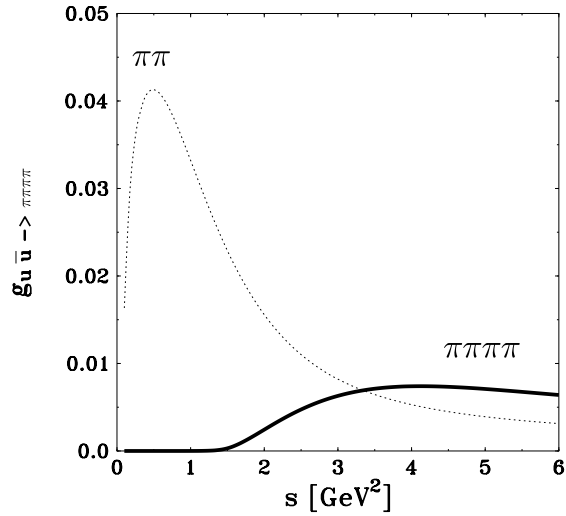,width=9cm}}
 \caption{Contribution of the four-meson weights as function of $s$.
In the figure the weight $g_{u\bar u  \to \pi^0  \pi^0  \pi^0  \pi^0}(s)$
is displayed (solid line) in comparison to the two-meson
weight  $g_{u\bar u  \to \pi^+  \pi^-}$ (dotted line).}
\label{gupipipipi} 
\end{figure}

For the four-meson contribution we take into account only pions because
they are the lightest particles. Then the following combinations
contribute:
\begin{equation}
\pi^+\pi^-\pi^+\pi^-, \quad \pi^+\pi^-\pi^0\pi^0, \quad  \pi^0\pi^0\pi^0\pi^0\;.
\end{equation}
Because of the small mass difference the weights of the three combinations
are practically indistinguishable, so it is sufficient to calculate
one of them.
Again, we make use of the recurrence relation Eq.~(\ref{fourparticle}). 
Fig.~\ref{gupipipipi} shows the weight for the case $\pi^0\pi^0\pi^0\pi^0$,
but it is indistinguishable from any other possible combination with
charged pions. It is seen that the four-meson weight numerically only
becomes relevant for $s> 2\;{\rm  GeV}^2$. 
Therefore, no four-meson weights or even higher weights are 
taken into account in our calculation. 
\subsection{The total weight function and the $q\bar q \to \pi^+\pi^-$
fragmentation function}
\begin{figure}
\centerline{\psfig{figure=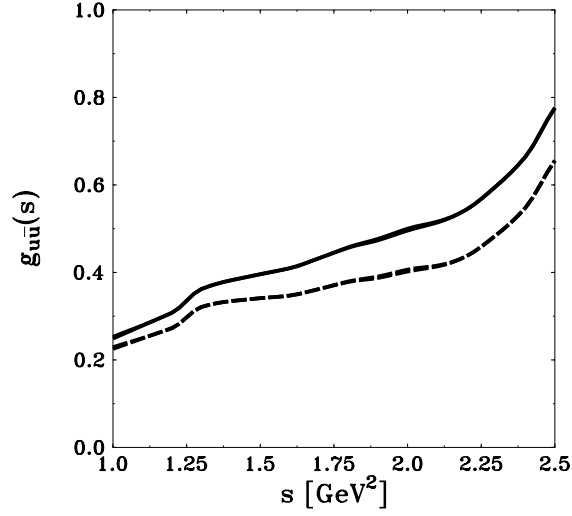,width=9cm}}
\caption{The total weight in the small-$s$ region.
The solid line shows the function $g_{u\bar u/ d\bar d}(s)$ as sum over all 
two- and three-meson  weights. The sum over
the two-meson  weights $g_{u\bar u / d\bar d \to 2}$ alone is
shown in the dashed line. One observes that the correction
from the three-meson weights is substantial.}
\label{gtot} 
\end{figure}

\begin{figure}
\centerline{\psfig{figure=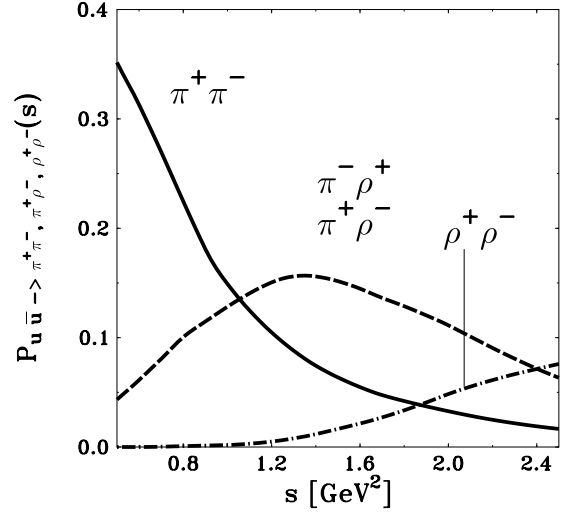,width=9cm}}
\caption{The fragmentation probability $P_{u\bar u \to A}$
as function of $s$. The
solid line shows the fragmentation probability for $A=\pi^+\pi^-$,
the dashed line for $A=\rho^+\pi^-$ or $A=\rho^-\pi^+$, and the
dashed-dotted line for $A=\rho^+\rho^-$.}
\label{pqqx} 
\end{figure}

Now we can compute the total weight function relevant in the region 
$s<2\;{\rm  GeV}^2$ or with some reservations 
also to  $s<2.5\;{\rm  GeV}^2$.
Fig.~\ref{gtot} shows the total weight $g_{u\bar u}(s)$ $\approx$
$g^{(+)}_{u\bar u \to 2}(s)$ +  $g^{(+)}_{u\bar u \to 3}(s)$
for the $\gamma^*\gamma$ reaction. 
The weight
$g_{d\bar d}(s)$ cannot be distinguished from $g_{u\bar u}(s)$
as the mass difference between charged and neutral kaons is negligible.
One observes that in the 
region $1\;{\rm GeV}^2<s<2.5\;{\rm GeV}^2$ the correction
of the three-meson weights is substantial. The same
observations hold also for the total $C=-$ weight, which is
not displayed here.
\newline
\newline
Fig.~\ref{pqqx} shows the $\gamma^* \gamma$ fragmentation probabilities 
$P_{q\bar q \to \pi\pi}(s)$, $P_{q\bar q \to \pi\rho}(s)$, 
and $P_{q\bar q \to \rho\rho}(s)$, where $q$ can be an $u$ or a $d$ quark. 
One observes an effective suppression for the production of $\rho$ mesons.
The reason lies in the fact that
the $\rho$ resonance appears later in $s$, where there
is more competition with other channels.
One should bear in mind that the primary production ratio between pseudoscalar
and vector mesons in the model is 1:1.
\subsection{Comparison to the time-like pion form factor}
\label{secpiform}
\begin{figure}
\centerline{\psfig{figure=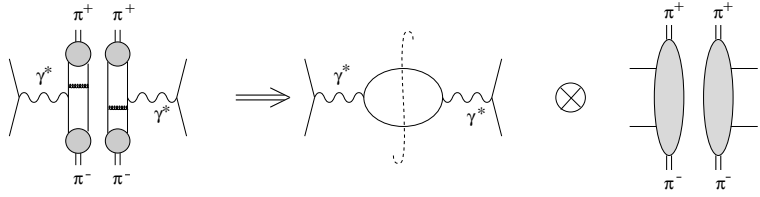,width=10cm}}
 \caption{Extraction of the fragmentation probability $P_{q \bar q \to \pi^+
\pi^-}(s)$ from the time-like pion form factor.}
\label{piformex} 
\end{figure}

It is interesting to confront our mechanism
with data one knows from the time-like pion form factor $F_\pi(s)$. In
the region  $1\;{\rm GeV}^2 < s < 4\; {\rm GeV}^2$ one has data
from Novosibirsk \cite{Barkov:1985ac} and the DM2 Collaboration
\cite{Bisello:1989hq}. Here the pion form factor is measured in
the $e^+ e^-$ annihilation process where the cross section is given
let alone from mass corrections by
 (see  \cite{Achasov:1998it,Akhmetshin:1999uj}):
\begin{equation}
\sigma(e^+ e^- \to \pi^+\pi^-)(s) 
= \frac{\pi \alpha_{\rm em}^2}{3s}|F_\pi(s)|^2\;.
\end{equation}
In our approach, this should be the same as the cross section for the
annihilation reaction $e^+e^- \to q\bar q$ times the subsequent 
fragmentation
probability into the $\pi^+ \pi^-$ pair (see Fig.~\ref{piformex}):
\begin{equation}
\sigma(e^+ e^- \to \pi^+\pi^-)(s) =\sum_q \sigma(e^+ e^- \to q\bar q)(s) 
P^{(-)}_{q\bar q \to \pi^+ \pi^-}(s) = 
\sum_q \frac{4 \pi \alpha_{\rm em}^2}{s} e_q^2
 \frac{g_{q\bar q \to \pi^+ \pi^-}(s)}{g^{(-)}_{q\bar q}(s)}\;.
\end{equation}
In contrast to the production of two pions from $\gamma^* \gamma$, the 
final state in $e^+ e^-$ annihilation must have the quantum number
$C=-$. This means that we have to take a total weight function 
${g^{(-)}_{q\bar q}(s)}$ which includes all relevant $C=-$ states. 
Our model will be a good description for the decay of 
resonances with masses larger than 1 GeV, where the resonance can
be interpreted as a string. In order to compare with data we have to 
include the fact that because of the final state being of $C=-$ a 
considerable contribution comes from the decay of one $\rho(770)$ meson
(neglecting the G parity violating transitions $\omega,\Phi \to \pi \pi$).
We can take  into consideration this contribution 
via the Vector Meson Dominance model
(VMD) \cite{Gounaris:1968mw}:
\begin{equation}
|F_\pi(s)|^2_{\rm VMD} = \frac
{g_{\rho \gamma}^2g_{\rho \pi \pi}^2}
{(s-m_\rho^2)^2 + \frac{m_\rho^6 \Gamma_\rho^2}{s^2}
\left( \frac{s-4 m_\pi^2}{m_\rho^2 - 4 m_\pi^2} \right)^3}\;.
\end{equation}
We take the value $g_{\rho \gamma}g_{\rho \pi \pi} = 0.705$
\cite{Aguilar-Benitez:1986fu,Perez-Y-Jorba:1977rh}. In Fig.~\ref{pifofit} 
it is shown that
adding the string fragmentation contribution to the VMD contribution 
yields a qualitatively good semi-classical description of the 
time like pion form factor. Of course the incoherent ansatz cannot
model the interference structure of the resonances, but it gives
a good description on the average, exactly what this semi-classical
picture should be.
\subsection{The process $\gamma^* \gamma \to \pi^+\pi^-$ at LEP2}
\begin{figure}
\centerline{\psfig{figure=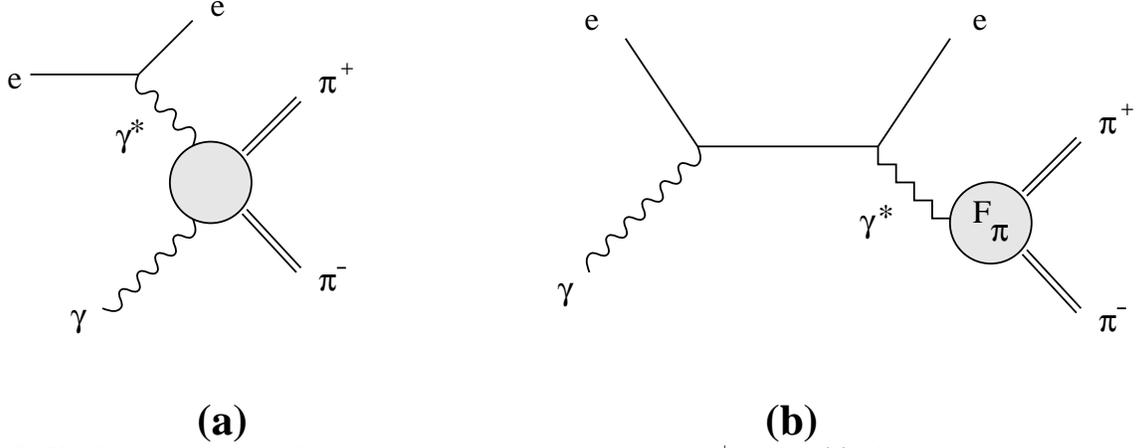,width=15.0cm}}
\caption{Subprocesses contributing to the reaction 
$e+ \gamma \to e +\pi^+ +\pi^-$: (a) pion pair production
via $\gamma^* \gamma$ scattering, (b) via bremsstrahlung.}
\label{brems}
\end{figure}

\begin{figure}
\centerline{\psfig{figure=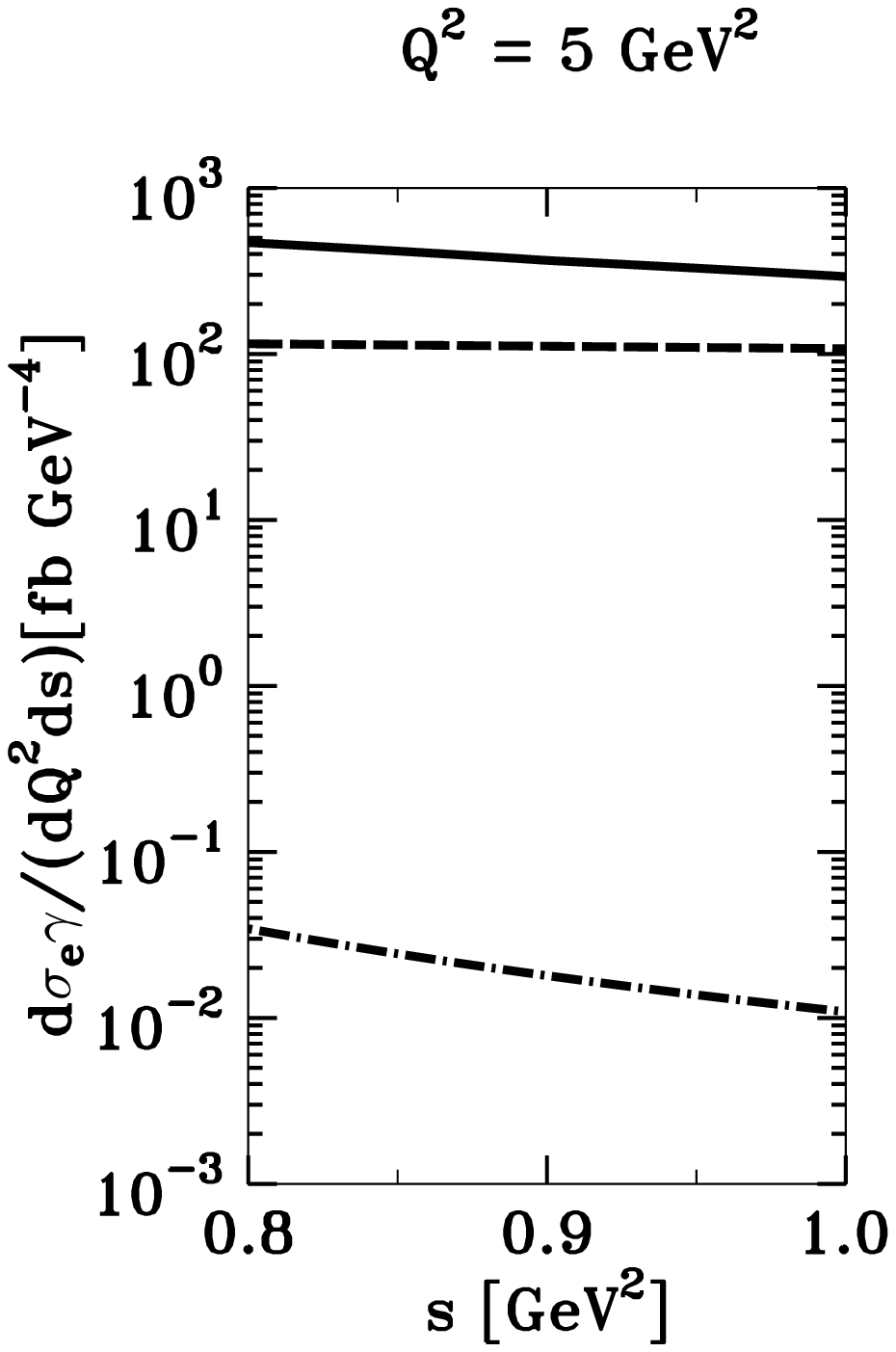,width=6.0cm}
            \psfig{figure=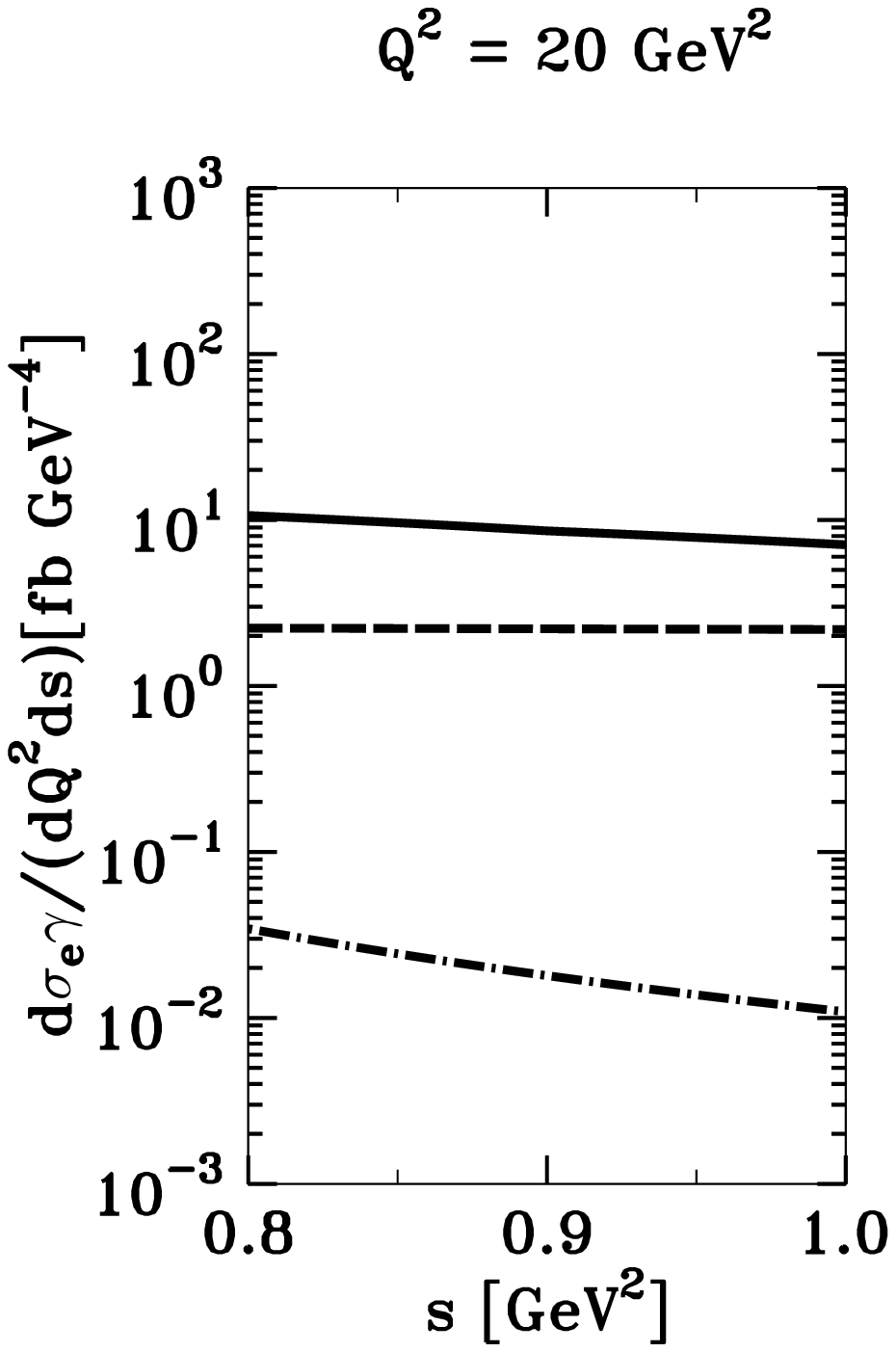,width=6.0cm}
            \psfig{figure=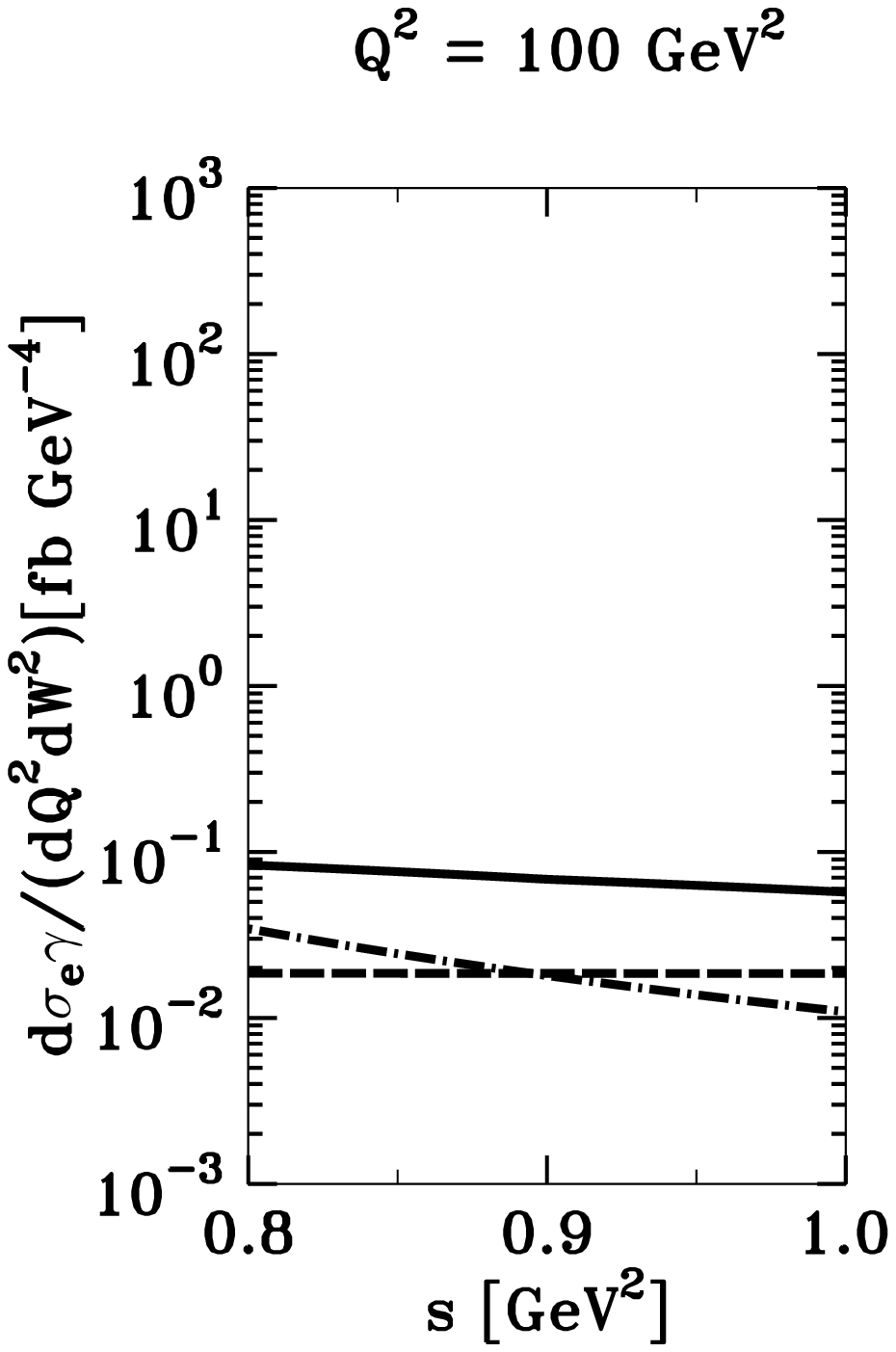,width=6.0cm}
}
\caption{Comparison of the Lund model calculation (solid line) with the model
given in Ref.~[11] (dashed line) for the cross section 
$d\sigma_{e\gamma}/(dQ^2 ds)$ for three different values of $Q^2$ as
function of $s$. The
dashed dotted line shows the contribution of the bremsstrahlung process as
given in Ref.~[11]. For the invariant mass of the electron photon system
we have taken in all figures the value $\sqrt{S_{e\gamma}}= 60 \; {\rm GeV}$.}
\label{xseccomp}
\end{figure}

Finally, 
we calculate the $\gamma^*\gamma \to \pi^+ \pi^-$ cross section. 
As a small digression we want to show how to reconstruct $p_\perp$
from Lorentz-invariant and experimentally accessible quantities.
Defining the variables $t = (p - q)^2$, $t'= (p-k)^2$, 
and $k=xq_0+q$ we get the relation:
\begin{eqnarray}
t' &=& m_\pi^2 - 2 pk 
\nonumber \\
   &=& m_\pi^2 - 2\left( \frac{1}{2} \sqrt{s}, \vec p_\perp, 
        \sqrt{ \frac{1}{4}- \frac{m_{\pi\perp}^2}{s} } \sqrt{s} \right)
           \left( \frac{1}{2} \sqrt{s}, \vec 0, \frac{1}{2} \sqrt{s}\right)
\nonumber \\
  &=& -(s z -  m_\pi^2 ) = (1-x)t + xm_\pi^2 \;.
\label{hit1}
\end{eqnarray}
Then for the total cross section one can in principle rewrite
the $p_\perp$ dependence as follows:
\begin{eqnarray}
\frac{ d\sigma( e e \to ee \pi^+ \pi^-)}{dy_0 ds dQ^2_0 dQ^2 d^2 {\bf p_\perp}} 
&=& \sum_q e_q^2 
\frac{2\pi \alpha_{\rm em}^2}{ Q^6} \frac{(1+(1-y)^2)}
{\left(1+ s/Q^2\right)^2} 
f_q^\gamma(x,Q^2,Q^2_0) f^T_{\gamma/e}(y_0, Q^2_0) 
P_{q\bar q \to \pi^+ \pi^-}(s,p_\perp^2)
\nonumber \\
\nonumber \\
\frac{ d\sigma( e e \to ee \pi^+ \pi^-)}{dy_0 ds dQ^2_0 dQ^2 dt} 
&=& \sum_q e_q^2 
\frac{2\pi \alpha_{\rm em}^2}{ Q^6} (1+(1-y)^2)
f_q^\gamma(x,Q^2,Q^2_0) f^T_{\gamma/e}(y_0, Q^2_0) 
\nonumber \\
&& \times 
\pi\frac{s/Q^2}{(1+s/Q^2)^3}\sqrt{1- 4\frac{m_{\pi\perp}^2}{s} } 
P_{q\bar q \to \pi^+ \pi^-}(s,p_\perp^2)
, \quad
p_\perp^2 = s z(1-z) - m_\pi^2\;.
\label{hit2}
\end{eqnarray}
So, $p_\perp^2$ can be reconstructed from $t$,$s$, and $Q^2$. 
Unfortunately, the luminosity that we will consider here
is not sufficient to trace the
$p_\perp$ dependence of the cross section. But we can instead have a look
at the $s$ dependence of the total $p_\perp$-integrated cross section.
For the numerical integration we use the standard LEP-2 parameters
\cite{Aurenche:1996mz} with a luminosity ${\cal L} = 500\;{\rm pb}^{-1}$ 
and the $e^+ e^-$ center of mass energy $\sqrt{S}= 175 \;{\rm GeV}$.
For $Q^2$ the allowed and measurable range is 
$5 \;{\rm GeV}^2 < Q^2 < 500 \;{\rm GeV}^2$. For the lower boundary of $Q^2$
we have to choose a rather low value in order to get a considerable rate,
although higher-twist contributions may be quite important here.
For the other cuts we choose:
\begin{eqnarray}
0.01 & < y_0 < & 0.99 \nonumber\\
1\; {\rm GeV}^2 & < s <& 2.5\; {\rm GeV}^2\;.
\end{eqnarray}
We approximate the total cross section with the $Q_0^2$ integrated
Weizs\"acker-Williams spectrum \cite{Frixione:1993yw}:
\begin{eqnarray}
f_{\gamma/e}^T(y,Q^2_{\rm WW})  &=& \frac{\alpha_{\rm em}}{2 \pi}
\left[ \frac{(1+(1-y)^2)}{y} \ln \frac{Q^2_{\rm WW}(1-y)}{m_e^2 y^2}
+ 2 m_e^2 y \left(\frac{1}{Q^2_{\rm WW}} - \frac{1-y}{m_e^2 y^2} 
\right)\right]\;.
\end{eqnarray}
We choose for the Weizs\"acker-Williams scale $Q^2_{\rm WW}= (m_1+m_2)^2$,
where $m_1$ and $m_2$ are the masses of the two mesons produced 
\cite{Frixione:1993yw,Frixione:1997ks}, hereby assuming that the maximum 
virtuality of the quasi real photon is smaller than 0.09 GeV$^2$. 
For the electromagnetic coupling constant
we choose a constant value $\alpha_{\rm em} = 1/137$.
Then, we get the approximated (and $p_\perp$-integrated) formula:
\begin{eqnarray}
\frac{ d\sigma( e e \to ee \pi^+ \pi^-)}{dy_0 ds  dQ^2} 
&\approx& \sum_q e_q^2 
\frac{2\pi \alpha_{\rm em}^2}{ Q^6} \frac{(1+(1-y)^2)}
{\left(1+s/Q^2\right)^2} 
f_q^\gamma(x,Q^2,0\;{\rm GeV}^2) f^T_{\gamma/e}(y_0, Q^2_{\rm WW}) 
P_{q\bar q \to \pi^+ \pi^-}(s)\;,
\end{eqnarray}
using again
\begin{equation}
P_{q\bar q \to \pi^+ \pi^-}(s) = 
\int d^2 {\bf p_\perp} P_{q\bar q \to \pi^+ \pi^-}(s,p_\perp^2)\;.
\end{equation}
For the photon structure function we choose the set SAS2 
($\rm \overline{MS}$ scheme)\cite{Schuler:1995fk,Schuler:1996fc}.
Fig.~\ref{xsec} shows the cross section for the LEP2 parameters above.
In the given kinematical range  the contribution of $\pi^+ \rho^-$ dominates
over the $\pi^+ \pi^-$ one. 
\newline
\newline
%
%
Experimentally, exclusive $\gamma^* \gamma$ scattering is a subprocess of
the more general exclusive $e\gamma$ scattering, see Fig.~\ref{brems}:
Taking the process  $e \gamma \to e \pi^+\pi^-$ for example, the 
measurable cross section consists of a contribution from
$\gamma^*\gamma$ scattering (Fig.~\ref{brems}(a)), 
and a background process of bremsstrahlung (Fig.~\ref{brems}(b)). The 
cross sections for both reactions have been estimated in \cite{Diehl:2000av}.
We can use their model predictions and calculations for the 
cross section $d\sigma_{e\gamma\to e \pi^+\pi^-}/(ds dQ^2)$
to compare with our
results of the Lund model, see Fig.~\ref{xseccomp}. Here we have plotted
the Lund model prediction for $\gamma^*\gamma$ scattering versus the
result for  $\gamma^*\gamma$ scattering and the bremsstrahlung background
from \cite{Diehl:2000av} for three different values
of $Q^2$, choosing for the invariant mass of the photon-electron system
$\sqrt{S_{e\gamma}} = 60 \;{\rm GeV}$. It turns out that at the matching
point $s$ = 1 GeV$^2$ the Lund model prediction is a factor 3.5 larger than
the model prediction in \cite{Diehl:2000av}. The discrepancy increases
to a factor of nearly 5 if one goes down to $s$ = 0.8 GeV$^2$, 
but at that value of $s$
it is doubtful whether the Lund model is reliable any longer. 
Considering that the model 
assumptions made in our case are still crude and that in  \cite{Diehl:2000av}
they estimate their model to be correct roughly by a factor of 2, the
discrepancy is tolerable, and the model predictions
shown here will be improved as soon as 
data are available. One can conclude, that the bremsstrahlung contribution
is also negligible for $Q^2 < 100 \;{\rm GeV}^2$ and $s > 1\;{\rm GeV}^2$. 
The interference contribution
between $\gamma^*\gamma$ scattering and bremsstrahlung vanishes if one
integrates over the azimuthal angle between the planes defined by the in 
and outgoing lepton on the one hand and the two produced pions on the other
hand, which we have done here.
%
%
\section{Summary and conclusions}
\label{sec4}
In this contribution we described the production of two-meson states
at intermediate momentum transfers in a semi-classical picture using
elements of the Lund model. In contrast to the well known application
of this model to high energy physics we counted all states explicitly
and took spin and $C$-parity into consideration. The model gives
a fairly good description at intermediate momentum transfers
above 1 GeV$^2$ when the decay of meson resonances
can be identified with the breaking of a string. This can be seen from
the fact that we get a consistent description for the time-like pion
form factor averaging over all interference effects. The procedure
has the potential to be the basis of a Monte Carlo program for 
intermediate energies. As an application we have used this picture
to predict the cross section $\gamma^*\gamma \to \pi^+\pi^-$, which is
interesting because it is sensitive to the  two-particle distribution
amplitude and  offers the possibility to observe the decay of a single string
and the formation of hadrons from quarks. The cross section
should be sizable at LEP2. 
\newline
\newline
{\bf Acknowledgment:} The author acknowledges stimulating discussion
with B.~Andersson, J.~Bijnens, G.~Gustafson, L.~L\"onnblad and T.~Sj\"ostrand,
and especially with M.~Diehl, who also sent the program for the comparison of his model
with the calculations performed in the Lund model.

\newpage

\newpage


\newpage


\newpage


\newpage

\newpage

\newpage

\newpage

\newpage




\newpage
\begin{table}
\begin{tabular}{|c||c|c|}
\hline && \\
 &  strangeness $S = 0$ mesons  & strangeness $S=1$ mesons \\ 
&& \\
\hline \hline
&& \\
charged &  $\pi^+ \pi^-$; $\rho^+ \rho^-$ & $K^+K^-$; $K^{*+} K^{*-}$ \\
        &  $\pi^+ \rho^-$; $\rho^+\pi^-$  & $K^{*+}K^-$; $ K^+K^{*-}$ \\
&& \\
\hline
&& \\
neutral & all $C=+$-meson combinations: $\pi^0$, $\eta$, $\eta'$ 
            &                         $K^0\bar K^0$; $K^{*0} \bar K^{*0}$ \\
            & all $C=-$-meson combinations: $\rho^0$, $\omega$
            &                         $K^{*0}\bar K^0$; $ K^0 \bar K^{*0}$ \\
&& \\
\hline
\end{tabular}
\caption{The possible  combinations for exclusive two-meson production in 
$\gamma^* \gamma$ scattering, if the primary quark-antiquark pair is $d\bar d$
or $u\bar u$.}
\label{allowedpairs}
\end{table}

\begin{table}
\begin{tabular}{|c||c|c|}
\hline && \\
particle & mass m [GeV] & width $\Gamma$ [GeV]\\
&& \\ \hline \hline
&& \\
$\pi^\pm$      & 0.13957 & -- \\
$\pi^0 $       & 0.13498 & -- \\ && \\
$\rho^\pm $    & 0.76690 & 0.14900 \\
$\rho^0   $    & 0.76850 & 0.15100 \\ && \\
$K^\pm$        & 0.49360 &  -- \\
$K^0$          & 0.49767 &  -- \\
$K^{*\pm}$     & 0.89160 & 0.04980 \\
$K^{*0}$       &0.89160  & 0.05050 \\  && \\
$\eta$ &0.54745 &-- \\
$\eta'$ &0.95777 & --\\
$\omega$ &0.78194 & 0.00843  \\
&& \\
\hline
\end{tabular}
\caption{Masses and widths used for calculating $g_{u \bar u}(s)$. 
In case no number is given for the width, it has been neglected 
in the calculation.}
\label{widthmass}
\end{table}
\newpage

\end{document}